\documentclass[11pt]{article}
\pdfoutput=1
\usepackage{jcappub}

\usepackage{color}
\usepackage{amsfonts, amsmath}
\usepackage{booktabs}
\usepackage[english]{babel}
\usepackage{colortbl}
\usepackage{epsfig}
\usepackage{epstopdf}
\usepackage{graphicx}
\usepackage{subfig}
\usepackage{lipsum}
\usepackage{hyperref}

\usepackage[load-configurations=astronomy, range-units=brackets, range-phrase=-, per-mode=reciprocal, mode=math]{siunitx}

\newcommand{\eq}[1]{Eq.~(\ref{#1})}
\newcommand{\vbunit}[1]{\hat{\vb{#1}}}
\newcommand{\convolved}[1]{\tilde{#1}}

\newcommand{\fig}[1]{Figure~\ref{#1}}

\newcommand{\vb}[1]{\mathbf{#1}}
\def\ie{{\em i.e.}~}
\def\eg{{{\em e.g.}~}}
\newcommand{\avg}[1]{\ensuremath{\left\langle \,#1\, \right\rangle}}

\newcommand{\tj}[6]{ \begin{pmatrix}
   #1 & #2 & #3 \\
   #4 & #5 & #6 
  \end{pmatrix}}

\def\Mpc{\, h^{-1} \, {\rm Mpc}}

\def\Gpc{\, h^{-1} \, {\rm Gpc}}

\def\kMpc{\, h \, {\rm Mpc}^{-1}}

\def\ie{{\em i.e.}~}
\def\eg{{{\em e.g.}~}}
	
\title{Interpreting measurements of the anisotropic galaxy power spectrum}

	\author[a,b]{Florian Beutler}
    \author[c,b]{Emanuele Castorina}
    \author[d]{Pierre Zhang} 
    \emailAdd{florian.beutler@port.ac.uk}
    \emailAdd{ecastorina@berkeley.edu}
    \emailAdd{pxyzhang@mail.ustc.edu.cn}
	
	\affiliation[a]{Institute of Cosmology \& Gravitation, University of Portsmouth, Portsmouth, PO1 3FX, UK}
		\affiliation[b]{Lawrence Berkeley National Laboratory, 1 Cyclotron Road, Berkeley, CA 94720, USA}
		\affiliation[c]{Berkeley Center for Cosmological Physics, Department of Physics, University of California, Berkeley, CA 94720}
        \affiliation[d]{CAS Key Laboratory for Researches in Galaxies and Cosmology, Department of Astronomy,
University of Science and Technology of China, Hefei, Anhui 230026, China
}

\abstract{
	The most commonly used estimators of the anisotropic galaxy power spectrum employ Fast Fourier transforms, and rely on a specific choice of the line-of-sight that breaks the symmetry between the galaxy pair. This leads to wide-angle effects, including the presence of odd power spectrum multipoles like the dipole ($\ell = 1$) and octopole ($\ell = 3$). In Fourier-space these wide-angle effects also couple to the survey window function. We present a self-consistent framework extending the commonly used window function treatment to include the wide-angle effects. We show that our framework can successfully model the wide-angle effects in the BOSS DR12 dataset. We present estimators for the odd power spectrum multipoles and, detect these multipoles in BOSS DR12 with high significance. Understanding the impact of the wide-angle effects on the power spectrum multipoles is essential for many cosmological observables like primordial non-Gaussianity and the detection of General Relativistic effects and represents a potential systematic for measurements of Baryon Acoustic Oscillations and redshift-space distortions.}
    
\begin{document}
\maketitle

\section{Introduction}

Measurements of $n$-point functions are a standard analysis tool in galaxy redshift surveys, since they, in principle, allow to extract all cosmological information even for non-linear point distributions. However, many of the symmetries which can simplify the work with point distributions are not present in galaxy redshift surveys.

Redshift space distortions (RSD), \ie the contribution of peculiar velocities to the measured redshift of a galaxy, are well known to break homogeneity and isotropy of $n$-point functions. The presence of a survey mask and of a redshift dependent galaxy selection function also breaks translational and rotational invariance, although with a different pattern than RSD. The importance of these effects depends on the definition of the line of sight (LOS) between the location of the observer and the object.  
The goal of this work is to clarify the interpretation of measurements of the galaxy power spectrum multipoles, including all the important physical effects in a self-consistent way.

The galaxy over-density field $\delta_g$ can be used to estimate the $2$-point function in Fourier-space using the so called Yamamoto estimator~\citep{Yamamoto2005:astro-ph/0505115v2}
\begin{equation}
P_{\ell}(k) = \langle \hat{P}_{\ell}(k) \rangle = \left\langle\frac{(2\ell + 1)}{2A}\int d\vb{s}_1 \int d\vb{s}_2\,\delta_g(\vb{s}_1)\delta_g(\vb{s}_2)e^{i\vb{k}\cdot(\vb{s}_1-\vb{s}_2)}\mathcal{L}_{\ell}(\vbunit{k}\cdot \vbunit{d}) - S_{\ell}\right\rangle,
\label{eq:symestimator}
\end{equation}
where $\vb{s}_{1,2}$ are redshift space galaxy positions and $\mathcal{L}_\ell$ is a Legendre polynomial. The normalisation and the shot noise term are given by $A = \int d\vb{s}\,n^2_g(\vb{s})$ and $S_\ell = (1+\alpha)\int d\vb{s}\,n_g(\vb{s})\mathcal{L}_{\ell}(\hat{\vb{k}}\cdot\hat{\vb{s}})$, respectively, where $\alpha$ is the fractional size of the random catalog compared to the data catalog.
The line-of-sight (LOS) direction $\hat{\vb{d}}$ is defined for each galaxy pair. The choice of the LOS is crucial, since any anisotropic signal is defined with respect to this LOS.
The monopole, $\ell=0$, of the Yamamoto estimator corresponds to the well known estimator of Feldman, Kaiser and Peacock (FKP)~\citep{Feldman1993:astro-ph/9304022v1}.

\begin{figure}[t]
\centering
\includegraphics[width=1.\textwidth]{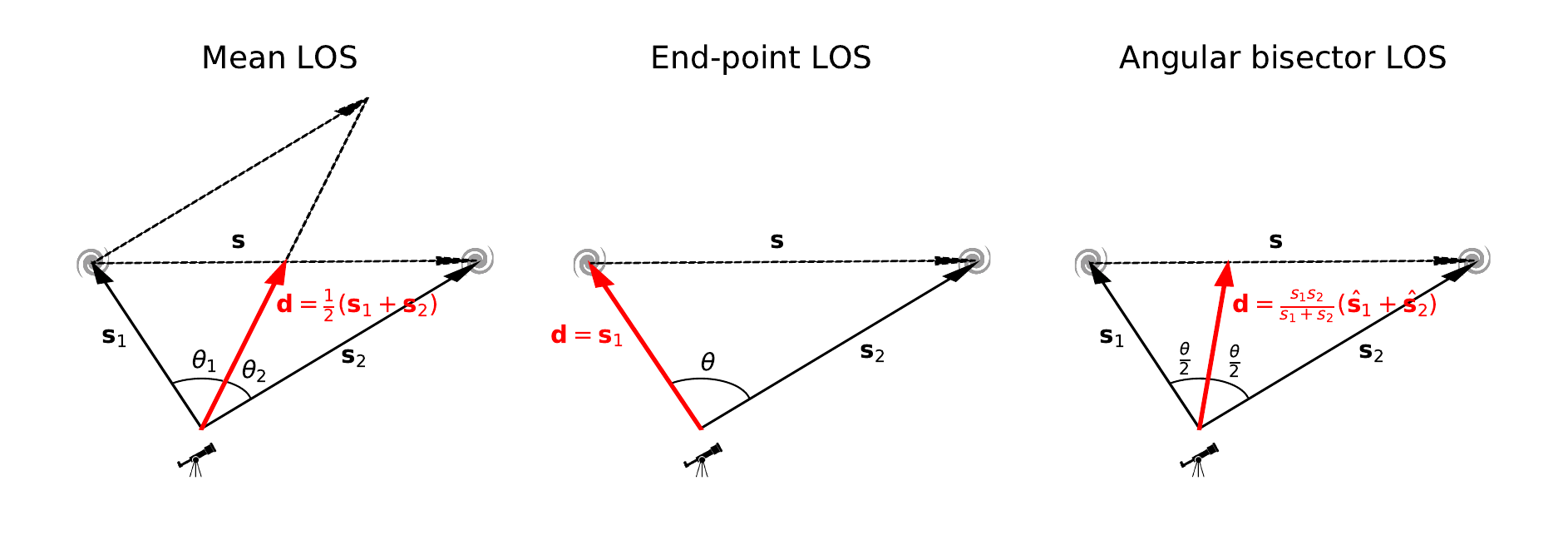}
\caption{Comparison of the line-of-sight (LOS) definitions used in this paper. The black arrows mark the triangle formed by the two galaxies and the observer. The red arrows show three possible choices for the LOS. The left diagram shows the mean LOS definition, the diagram in the middle shows the end-point LOS definition, used by the FFT-based power spectrum multipole estimators and the diagram on the right shows the angular bisector LOS. The mean and angular bisector LOS definitions keep the symmetry between the galaxy pair, which ensure the odd multipoles to be zero, which is not true for the end-point LOS.}
\label{fig:wideangle}
\end{figure}

A reasonable choice for the LOS is the mean vector $\vb{d} = \vb{s}_h = \frac{1}{2} (\vb{s}_1 + \vb{s}_2)$ (see \fig{fig:wideangle}), which has the advantage of being symmetric in the pair members, meaning the exchange of the two galaxies would lead to the same power spectrum. This property enforces the odd power spectrum multipoles to be zero.

However, the choice of  $\vb{s}_h$ as LOS does require  recalculating that quantity for each galaxy pair, which naturally leads to an $\mathcal{O}(N^2)$ algorithm, where $N$ indicates the size of the dataset. 
Modern datasets have $N \sim 1\,000\,000$ (see~\citep{Reid2015:1509.06529v2}), while future datasets like DESI~\citep{Aghamousa2016:1611.00036v2} and Euclid~\citep{Amendola2012:1206.1225v2} will have $N \sim 50\,000\,000$. The size of these future datasets represents a computational challenge, even for the  calculation of a simple $2$-point statistic.
Moreover, the analysis of galaxy surveys heavily relies on mock datasets to test the clustering models and generate covariance matrices. This usually means that beside the dataset itself, $\gtrsim 1000$ mock datasets need to be analyzed in parallel.
Practical considerations therefore suggest to look for faster FFT-based estimators. 
A simple FFT-based estimator can be constructed from \eq{eq:symestimator} if one defines the LOS as $\vbunit{d} = \vbunit{s}_1$~\citep{Feldman1993:astro-ph/9304022v1,Bianchi2015:1505.05341v2,Scoccimarro2015:1506.02729v2,Slepian2015:1506.04746v2,Sugiyama2017:1704.02868v2,Hand2017:1704.02357v1,Slepian2017:1709.10150v1,Sugiyama2018:1803.02132v1}. From now on we will call this LOS definition the end-point LOS.

This choice of LOS has two main effects in a power spectrum analysis. First, wide-angle effects~\citep{Szalay1997:astro-ph/9712007v1,Sza04,PapSza08,Raccanelli2010:1006.1652v1,Raccanelli,Bertacca,Yoo2015,Rei16,Tansella2017,Castorina2017:1709.09730v2,Castorina2018:1803.08185v2}, are significantly larger in the end-point LOS definition compared to the mean LOS definition~\cite{Castorina2017:1709.09730v2,Castorina2018:1803.08185v2,Rei16}, and second it breaks the symmetry between the galaxy pair and therefore introduces non-zero odd power spectrum multipoles, like the dipole, $P_1(k)$ and the octopole, $P_3(k)$ \cite{Raccanelli,Bonvin14,Rei16,Castorina2018:1803.08185v2}.

Averaging wide-angle effects over the moving LOS in \eq{eq:symestimator}, will generate corrections to the well known plane-parallel formula of Kaiser \cite{Kaiser} that scale as $k^{-2}$. The latter are partially degenerate with primordial non-Gaussian signatures in the galaxy bias~\citep{Dalal,Jeong,Yoo2012,Camera} and including them in a cosmological analysis is crucial for an unbiased measurement of primordial non-Gaussianity. 

The odd power spectrum multipoles introduced by the end-point LOS definition are of geometric nature and as such they do not carry any extra cosmological information, but nonetheless they represent a possible contamination in the detection of physical dipoles generated by relativistic effects~\cite{Raccanelli,Bertacca,Bonvin14,Gaztanaga2015:1512.03918v2,Lepori2017:1709.03523v2}.
As we will show in Section~\ref{sec:wideangle}, the odd power spectrum multipoles couple to the even multipoles through the survey window function, and must be included in a self-consistent treatment. Both wide-angle effects and the presence of odd multipoles have so far been ignored when analyzing power spectrum measurements~\citep{Beutler2016:1607.03150v1,Beutler2016:1607.03149v1,Gil-Marin2016:1606.00439v2,Grieb}.  

Recently~\citep{Castorina2017:1709.09730v2,Castorina2018:1803.08185v2} proposed a simple formalism to compute the mean of the estimator in \eq{eq:symestimator} for a generic choice of the LOS, and discussed how to include various observational effects.
Starting from the aforementioned papers, the goal of this work is to present a self-consistent framework to interpret measurements of the multipoles of the galaxy power spectrum at large scales.
We include wide-angle effects and their possible coupling with the survey geometry, which accounts for the angular mask and the radial selection function. Our analysis further simplifies the treatment in~\citep{Castorina2017:1709.09730v2}.
We analyze the power spectrum multipoles of the Baryon Oscillation Spectroscopic Survey (BOSS), where we can clearly detect these effects. We find that in BOSS DR12, wide-angle effects introduced by FFT-based estimators and the standard window function effects are of the same order and can impact cosmological parameter constraints. We show that our formalism is able to capture all of these effects without introducing any new free parameters. 

The paper is organized as follows: In Section~\ref{sec:wideangle} we discuss our formalism to include the wide-angle effects in the modeling of the power spectrum multipoles. We then introduce the BOSS DR12 dataset in Section~\ref{sec:datamocks} and calculate the individual correction terms required for this sample. We conclude in Section~\ref{sec:conclusion}.

The fiducial cosmology for the mock data analysis (Patchy mock catalogs) follows the cosmology of the Patchy simulations, which is $\Lambda$CDM with $\Omega_m = 0.307115$, $\Omega_b=0.048206$, $\sigma_8 = 0.8288$, $n_s = 0.9611$ and $h=0.6777$. The BOSS data analysis follows the fiducial BOSS cosmology, which is $\Lambda$CDM with $\Omega_m = 0.31$, $\Omega_bh^2 = 0.022$, $h = 0.676$, $\sigma_8 = 0.8$, $n_s = 0.96$ and $\sum m_{\nu} = 0.06\,$eV.

\section{Modeling wide-angle effects}
\label{sec:wideangle}

In this section we discuss the wide-angle effects present in the 2-point statistics and how these terms depend on the definition of the line-of-sight. We start  with the more intuitive configuration-space picture before moving to Fourier-space.
More technical details can be found in~\citep{Castorina2017:1709.09730v2,Castorina2018:1803.08185v2}.

\subsection{The configuration-space picture}

Neglecting redshift space distortions, the 2-point correlation function  is statistically isotropic and homogeneous, and therefore can only depend on the separation between the two galaxies $s = |\vb{s}_1 - \vb{s}_2|$, which in configuration-space yields
\begin{equation}
\xi(s) = \langle \delta(\vb{s}_1)\delta(\vb{s}_2)\rangle.
\end{equation}
However, RSD  make the location of the observer a special place and therefore break isotropy and homogeneity. To describe the system we therefore need to fully specify the triangle between the two galaxies and the observer, see \fig{fig:wideangle}, and following \cite{Szalay1997:astro-ph/9712007v1} we write 
\begin{align}
\label{eq:xi}
\avg{\delta(\vb{s}_1)\delta(\vb{s}_2)}\equiv\xi(\vb{s}_1,\vb{s}_2) = \xi(\vb{s},\vb{d}) = \sum_{\ell_1,\ell_2} C_{\ell_1,\ell_2}(s) \mathcal{L}_{\ell_1}(\cos \theta/2)\mathcal{L}_{\ell_2}(\mu)
\end{align}
where $\cos\theta = \vbunit{s}_1\cdot \vbunit{s}_2$ is the cosine of the angle between the two galaxies as seen by the observer, and $\mu \equiv \hat{\vb{s}}\cdot \hat{\vb{d}}$. The coefficients $C_{\ell_1,\ell_2}(s)$ can be computed in perturbation theory~\citep{Szalay1997:astro-ph/9712007v1,Sza04,Castorina2017:1709.09730v2}. In the limit $\theta\rightarrow0$, \ie the opening angle between the two galaxies is very small, one recovers the familiar plane-parallel expressions for multipoles of the correlation function as derived for the first time by Kaiser~\citep{Kaiser}. In the following we will give a more rigorous definition of the plane-parallel limit as the lowest order series expansion in the wide-angle corrections.

The explicit dependence of \eq{eq:xi} on the opening angle $\theta$, and hence on the definition of the line of sight, encapsulates the wide-angle effects we focus on in this work. 
Formally we can expand the correlation function in the parameter $x_s=s/d$, where $d = |\vb{d}|$, and rewrite \eq{eq:xi} as
\begin{align}
\label{eq:xisd}
\xi(\vb{s},\vb{d}) = \xi(s,d,\mu) = \sum_\ell \xi_\ell(s,d)\mathcal{L}_\ell(\mu) = \sum_\ell \sum_n x_s^n \xi_\ell^{(n)}(s)\mathcal{L}_\ell(\mu).
\end{align}
The $n=0$ terms correspond to the plane-parallel multipoles, and
$n>0$ to wide-angle effects. The functions $\xi_\ell^{(n)}(s)$ depend on the choice of the LOS.
Usually the analyses of galaxy redshift surveys only include the $n=0$ terms~\citep{Beutler2013:1312.4611v2,Beutler2016:1607.03149v1}, which is referred to as the ``local" plane-parallel approximation.

\fig{fig:wideangle} shows three possible definitions of the LOS. The diagram on the left shows, in red, the mean LOS given by $\vb{d} = \frac{1}{2}(\vb{s}_1 + \vb{s}_2)$ and the diagram on the right shows the angular bisector LOS, $\vb{d} = \frac{s_1s_2}{s_1 + s_2}(\vbunit{s}_1 + \vbunit{s}_2)$. These two LOS definitions have the advantage that they are symmetric under the exchange of $\vb{s}_1$ and $\vb{s}_2$, which ensures that $\xi^{(n)}_{\ell} = 0$ for all odd $\ell$ and $n$. There are still wide-angle effects present for the mean and angular bisector LOS, but only at second order ($n=2$). 

The now standard FFT-based algorithms used to measure the power spectrum multipoles of \eq{eq:symestimator} rely on the end-point LOS definition ($\vb{d} = \vb{s}_1$) shown in the middle of \fig{fig:wideangle}. In this case we will see the first order wide-angle corrections ($n=1$) do not vanish and we have to include non-zero odd multipoles.

\subsection{The Fourier-space picture}

On the curved sky, the redshift space mapping breaks translational invariance of the galaxy clustering, such that the power spectrum and the correlation function are no longer a Fourier transform pair. This is expected, as the correlation function now depends on the triangle configuration defined by the observer and the pair of galaxies. It is therefore compelling to understand what the ensemble average of the power spectrum estimators in \eq{eq:symestimator} really yields, and how it compares to the plane-parallel limit employed in the data analyses~\citep{Samushia15}.
We start with a summary of the results in~\citep{Castorina2017:1709.09730v2,Castorina2018:1803.08185v2} to which we refer the reader for further details.

By taking the Fourier transform of \eq{eq:xisd} with respect to the separation vector $\vb{s}$, we can define a 'local', or LOS-dependent, power spectrum~\citep{Rei16,Scoccimarro2015:1506.02729v2}
\begin{align}
 P(\vb{k},\vb{d}) \equiv \int \mathrm{d}^3 s
    \, \xi(\vb{s},\vb{d})e^{-i\vb{k}\cdot\vb{s}},
\end{align}
which admits the following expansion in multipoles and wide-angle contributions,
\begin{align}
  P(\vb{k},\vb{d}) &= \sum_{\ell} P_\ell(k,d)
  \mathcal{L}_\ell(\vbunit{k}\cdot\vbunit{d})
\label{eqn:Pkd_expansion} \\
 &\equiv \sum_{\ell,n} x_k^{n} P_\ell^{(n)}(k)\mathcal{L}_\ell(\vbunit{k}\cdot\vbunit{d}),
\label{eqn:Pln_bisector}
\end{align}
where the expansion parameter is now $x_{k} = (k d)^{-1}$.  Again, the $n=0$ terms correspond to the familiar plane-parallel multipoles, and, at each order in $n$, we can write
\begin{align}
 P_\ell^{(n)}(k) = 4\pi(-i)^\ell\int s^2\,\mathrm{d}s\ (ks)^{n}\,\xi_\ell^{(n)}(s)\,j_\ell(ks)
\label{eqn:Pln_bisector_hankel}
\end{align}
with the inverse relation
\begin{equation}
  \xi_\ell^{(n)}(s) = i^\ell\int\frac{k^2\,\mathrm{d}k}{2\pi^2}\ (ks)^{-n}P_\ell^{(n)}(k)\,j_\ell(ks).
\end{equation}
If we neglect finite volume effects, which will be discussed in the next section, it can be shown that the Yamamoto estimator in \eq{eq:symestimator} measures the LOS-average of \eq{eqn:Pkd_expansion}
\begin{align}
\avg{P_{\ell}(\vb{k})} = \frac{1}{V}\int \mathrm{d}^3 d \, P_\ell(k,d),
\end{align}
where $V$ is the survey volume. The above equation shows that the Yamamoto estimator can be thought of as the average across all possible line of sights of a local power spectrum, which has a well defined plane-parallel limit. The importance of wide-angle effects will depend on how much the chosen LOS varies within the survey.

\subsection{Wide-angle effects in linear perturbation theory}

As mentioned several times in the previous sections, the wide-angle correction terms depend on the definition of the LOS. Here we start with the angular bisector LOS definition, shown on the right in \fig{fig:wideangle}, before moving on to the case of the end-point LOS.
This topic has been discussed at length elsewhere~\citep{Szalay1997:astro-ph/9712007v1,Sza04,PapSza08,Raccanelli2010:1006.1652v1,Raccanelli,Bertacca,Yoo2015,Rei16,Tansella2017,Castorina2017:1709.09730v2,Castorina2018:1803.08185v2}, and here we only report the equations relevant for our work.
Since wide-angle effects are only important at very large scales, where $x_s=s/d$ differs from zero, we will model the $n>0$ corrections to \eq{eq:symestimator} using linear theory.
In the case of the angular bisector LOS,  the $n=1$ terms in \eq{eqn:Pln_bisector} vanish and hence the leading order wide-angle corrections at $n=2$ are \cite{Rei16,Castorina2017:1709.09730v2}:
\begin{align}
\label{eq:xiWA1}
  \xi_0^{{\rm bisector},(2)}(s) &= -\frac{4\beta^2}{45}\ b_1^2 \Xi_0^{(0)}(s) - \frac{\beta(9+\beta)}{45}\ b_1^2\Xi_2^{(0)}(s)\\
  \label{eq:xiWA2}
  \xi_2^{{\rm bisector},(2)}(s) &=  \frac{4\beta^2}{45}\ b_1^2\Xi_0^{(0)}(s) + \frac{\beta(189+53\beta)}{441}\ b_1^2\Xi_2^{(0)}(s) -\frac{4f^2}{245}\ b_1^2\Xi_4^{(0)}(s)\\
  \label{eq:xiWA3}
  \xi_4^{{\rm bisector},(2)}(s) &= -\frac{8\beta(7+3\beta)}{245}\ b_1^2\Xi_2^{(0)}(s) + \frac{4\beta^2}{245}\ b_1^2\Xi_4^{(0)}(s),
\end{align}
where $b_1$ is the linear bias, and we defined
\begin{equation}
  \Xi_\ell^{(n)}(s) = \int\frac{k^2\,\mathrm{d}k}{2\pi^2}\ (k)^{-n}P(k)\,j_\ell(ks)
  \label{eq:order2multi}
\end{equation}
with $P(k)$ being the linear, real space, matter power spectrum.

Now we move to the end-point LOS definition $\vbunit{d} = \vbunit{s}_1$. The drawback of this definition of the LOS is that the wide-angle effects are much larger compared to the mean or angular bisector LOS \cite{Rei16,Castorina2017:1709.09730v2}, and odd multipoles are generated at first order ($n=1$). 

The dipole ($\ell = 1$) and octopole ($\ell = 3$) in the end-point LOS are given by \cite{Bonvin14,Rei16,Castorina2017:1709.09730v2}
\begin{align}
\label{eq:order1term0}
&\xi_1^{{\rm ep},(1)}(s) =-\frac{3}{5}\xi_2^{(0)}(s), \\
	&\xi_3^{{\rm ep},(1)}(s) = \frac{3}{5}\xi_2^{(0)}(s)-\frac{10}{9}\xi_4^{(0)}(s)\,.
    \label{eq:order1term1}
\end{align}
As expected, given their geometric nature, odd multipoles do not carry any cosmological information, since they are proportional to the $n=0$ terms. 
Note that the above equations for the odd multipoles are valid at all orders in perturbation theory in the density fields and not just in linear theory.

The quadratic terms ($n=2$) for the end-point LOS definition are related to the equivalent terms of the angular bisector LOS (see Eq.~\ref{eq:xiWA1} to~\ref{eq:xiWA3}) and are given by
\begin{align}
  \xi_0^{{\rm ep},(2)}(s) &= \xi_0^{{\rm bisector},(2)}(s) + \frac{1}{5}\xi_2^{(0)}(s) \\
  \xi_2^{{\rm ep},(2)}(s) &=  \xi_2^{{\rm bisector},(2)}(s) -  \frac{2}{7}\xi_2^{(0)}(s)  + \frac{5}{7}\xi_4^{(0)}(s) \\
  \xi_4^{{\rm ep},(2)}(s) &=  \xi_4^{{\rm bisector},(2)}(s)+ \frac{3}{35}\xi_2^{(0)}(s) - \frac{90}{77}\xi_4^{(0)}(s).
  \label{eq:order2term}
\end{align}
We can now obtain the Fourier-space quantities using the generalized Hankel transforms defined in \eq{eqn:Pln_bisector_hankel}. Expressions for the wide-angle effects beyond linear theory, can be found in~\cite{Castorina2018:1803.08185v2} (using the Zeldovich approximation).

It is also well known that the galaxy selection function can generate additional wide-angle terms in the correlation function and power spectrum multipoles~\citep{Szalay1997:astro-ph/9712007v1, Yoo2015,PapSza08,Raccanelli2010:1006.1652v1,Castorina2017:1709.09730v2}. Analytic expressions for those can be found in Appendix~\ref{sec:selfunc}. The main difficulty in evaluating these wide-angle contributions is that they require knowledge of the underlying real space density of galaxies, which is usually not available. We therefore do not attempt to include them in our analysis, but rather argue in Appendix~\ref{sec:selfunc} that they would not qualitatively change our results.

\subsection{Accounting for the survey window function}
\label{sec:window}

The power spectrum we measure from a galaxy survey is related to the true underlying power spectrum by a convolution with the survey window function $W(\vb{s}_i)$. This convolution distorts the shape of the power spectrum, correlates modes and distributes power between the different multipoles, and therefore needs to be properly taken into account in the data analysis.

The standard approach to include the survey window is to (1) Hankel transform the underlying power spectrum model to configuration-space, (2) multiply it with the multipoles of the survey window function and (3) Hankel transform back into Fourier-space~\citep{Wilson2015:1511.07799v2, Beutler2016:1607.03150v1}. 

The convolved power spectrum estimated by a FFT-based estimator is given by 
\begin{align}
	\avg{\convolved{P}_A(\vb{k})} &= (2A+1)\int \frac{d \Omega_k}{4\pi} \,\mathrm{d}^3 s_1 \,\mathrm{d}^3 s_2 \,e^{-i \vb{k}\cdot(\vb{s}_1-\vb{s}_2)} \avg{\delta(\vb{s}_1) \delta(\vb{s}_2)}  W(\vb{s}_1)W(\vb{s}_2)\mathcal{L}_A(\vbunit{k}\cdot \vbunit{s}_1)\\
	\begin{split}
		& =(-i)^A (2A+1) \sum_{\ell,\, L}\tj{\ell}{L}{A}{0}{0}{0}^2(2L+1)\int \mathrm{d}s\,s^2 j_A (ks) \sum_n (s)^n\, \xi_\ell^{{\rm ep},(n)}(s) \\
		&\times \int \mathrm{d} \Omega_s \int \mathrm{d}^3 s_1  (s_1)^{-n} W(\vb{s}_1)W(\vb{s}+\vb{s}_1)\mathcal{L}_L(\vbunit{s}\cdot \vbunit{s}_1),
	\end{split}
    \label{eq:convolvedP}
\end{align} 
where $\convolved{P}$ is our notation for the power spectrum convolved with the survey window function. 

Defining the multipoles of the window functions with respect to the same LOS ($\vbunit{d} = \vbunit{s}_1$) gives 
\begin{align}
Q_L^{ep,(n)}(s) \equiv (2L+1) \int \mathrm{d} \Omega_s \int \mathrm{d}^3 s_1  (s_1)^{-n} W(\vb{s}_1)W(\vb{s}+\vb{s}_1)\mathcal{L}_L(\vbunit{s}\cdot \vbunit{s}_1)
\label{eq:Qnep}
\end{align}
resulting in
\begin{align}
\avg{\convolved{P}_A(\vb{k})}= (-i)^A (2A+1) \sum_{\ell,\, L}\tj{\ell}{L}{A}{0}{0}{0}^2\int \mathrm{d}s\, j_A (ks) \sum_n s^{n+2} \,\xi_\ell^{{\rm ep},(n)}(s) Q_L^{{\rm ep},(n)}(s).
\label{eq:convpower}
\end{align}
The above expression is exact at any order in the wide-angle corrections, and it represents one of our main results. 
It simplifies the formalism of~\citep{Castorina2017:1709.09730v2} in a way more suitable for power spectrum analysis. 
\eq{eq:convpower} contains the wide-angle corrections $n>0$ as well as the window function contributions. We will now move on to further investigate the implications of this equation, discussing each power spectrum multipole in turn, and the relation of our exact expression to various approximations present in the literature. 

\section{Case study: Power spectrum multipoles in BOSS DR12}
\label{sec:datamocks}

As a first application of our formalism we will now study the impact of wide-angle effects on measurements of power spectrum multipoles of the Data Release 12 (DR12) of the Baryon Oscillation Spectroscopic Survey (BOSS).

\subsection{The BOSS DR12 dataset}

The BOSS survey was part of SDSS-III~\citep{Eisenstein2011:1101.1529v2,Dawson2012:1208.0022v3} and used the SDSS multi-fibre spectrographs~\citep{Bolton2012:1207.7326v2,Smee2012:1208.2233v2} to measure spectroscopic redshifts of $1\,198\,006$ million galaxies, representing the currently largest dataset of its kind.
The galaxies were selected from multicolour SDSS imaging~\citep{1996AJ....111.1748F,Gunn1998:astro-ph/9809085v1,Smith2002:astro-ph/0201143v2,Gunn2006:astro-ph/0602326v1,Doi2010:1002.3701v1, Reid2015:1509.06529v2}
over $10\,252$ deg$^2$ divided in two patches on the sky called North Galactic Cap (NGC) and South Galactic Cap (SGC) and cover a redshift range of $0.2$ - $0.75$. In our analysis we split this redshift range into two redshift bins given by $0.2 < z < 0.5$ and $0.5 < z < 0.75$, which we from now on call z1 and z3, respectively. When analyzing the dataset we account for incompleteness following the standard BOSS analysis~\citep{Ross2012:1203.6499v3}.

\subsection{The Multidark Patchy mock catalogs}

The BOSS collaboration provides a set of mock catalogs, which mimic the clustering properties and geometry of the observed dataset~\citep{Kitaura2015:1509.06400v3}. These mock catalogues have been produced using approximate gravity solvers and analytical-statistical biasing models. The catalogues have been calibrated to a N-body based reference sample extracted from one of the BigMultiDark simulations~\citep{Klypin2014:1411.4001v2}, which was performed using Gadget-2~\citep{Springel2005:astro-ph/0505010v1} with $3\,840^{3}$ particles on a volume of [$2.5\Mpc$]$^3$ assuming a $\Lambda$CDM cosmology with $\Omega_m = 0.307115$, $\Omega_b = 0.048206$, $\sigma_8 = 0.8288$, $n_s = 0.9611$, and a Hubble constant of $H_0 = 67.77$\,km\,s$^{-1}$Mpc$^{-1}$.

Halo abundance matching is used to reproduce the
observed BOSS 2- and 3-point clustering measurements~\citep{Rodriguez-Torres2015:1509.06404v3}. This technique is applied at different redshift bins to reproduce the BOSS DR12 redshift evolution. These mock catalogues are combined into light cones, also accounting for the selection effects and survey mask of the BOSS survey. In total we have $2048$ mock catalogues available for each of the two patches of the BOSS dataset (NGC and SGC). 

There are two versions of these mock catalogues, named V6S and V6C. The V6C catalogues have been adjusted to better reproduce the observed power spectrum quadrupole and we use those catalogues to obtain the covariance matrix, while we use the V6S catalogues when testing our model of the power spectrum multipoles. The differences between these two versions of the mock catalogs are subtle and do not impact any of the conclusions of this analysis.

\subsection{The plane-parallel limit ($n=0$)}
\label{sec:planeparallel}

\begin{figure}[t]
\centering
\includegraphics[width=0.45\textwidth]{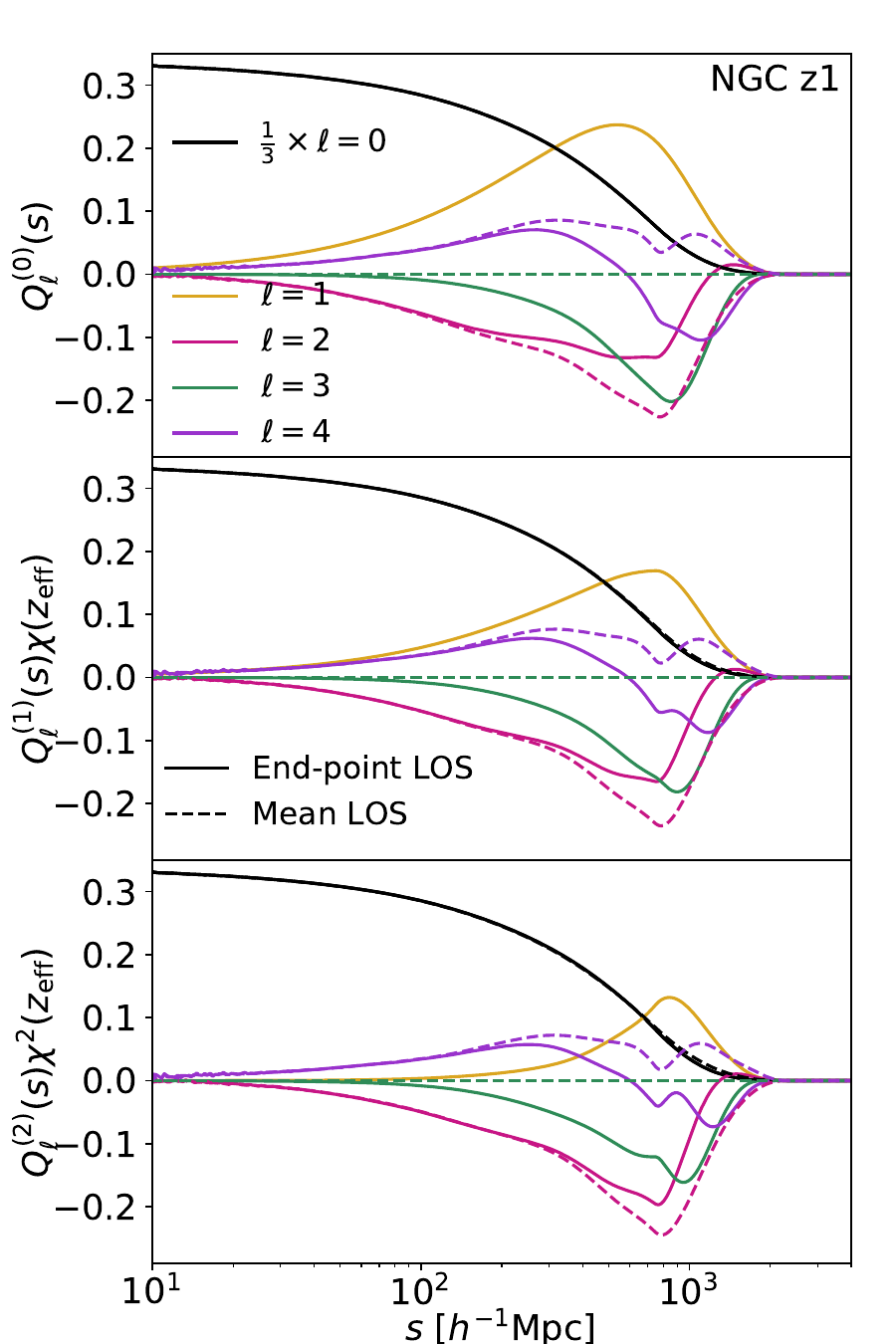}
\includegraphics[width=0.45\textwidth]{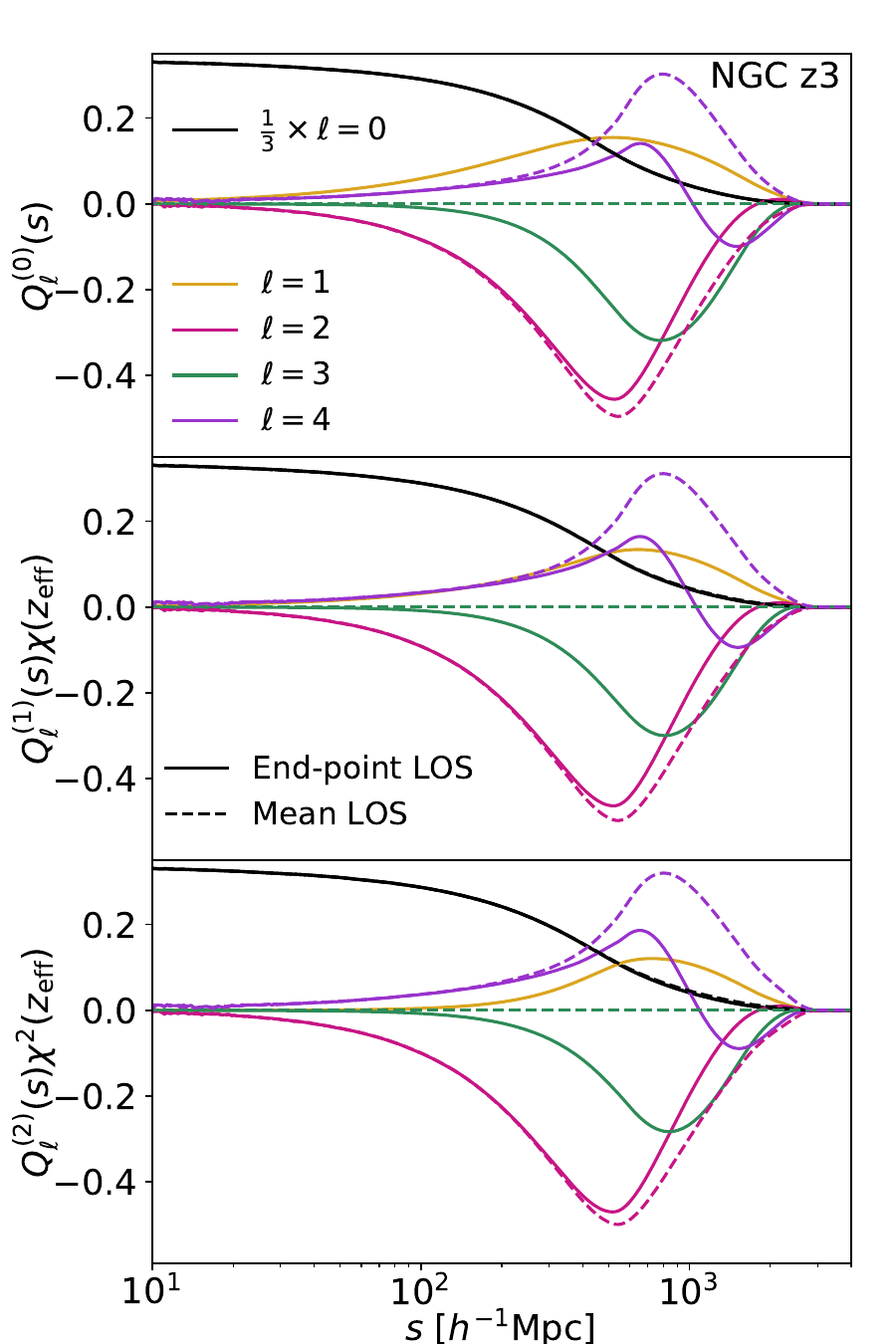}
\caption{Comparison of window function multipoles calculated by pair counting (rather than using the FFT-based estimator) using the mean LOS definition (dashed lines) and the end-point LOS definition (solid lines) for BOSS DR12, NGC. The corresponding plots for the SGC are shown in \fig{fig:cmpLOS_SGC}. The plot on the left shows the low redshift bin of BOSS ($0.2 < z < 0.5$, $z_{\rm eff} = 0.38$), while the plot on the right shows the high redshift bin ($0.5 < z < 0.75$, $z_{\rm eff} = 0.61$). The three panels show the window functions with different orders of the wide-angle expansion, $n$, where $\chi(z_{\rm eff}=0.38)=1034.8\Mpc$ and $\chi(z_{\rm eff}=0.61)=1560.5\Mpc$ represent the co-moving distances to the effective redshift of the low (z1) and high (z3) redshift bin, respectively. The monopole term is divided by 3 to improve the visibility of the higher order multipoles. The end-point LOS definition introduces non-zero odd multipoles.}
\label{fig:cmpLOS_NGC}
\end{figure}

\begin{figure}[t]
\centering
\includegraphics[width=0.45\textwidth]{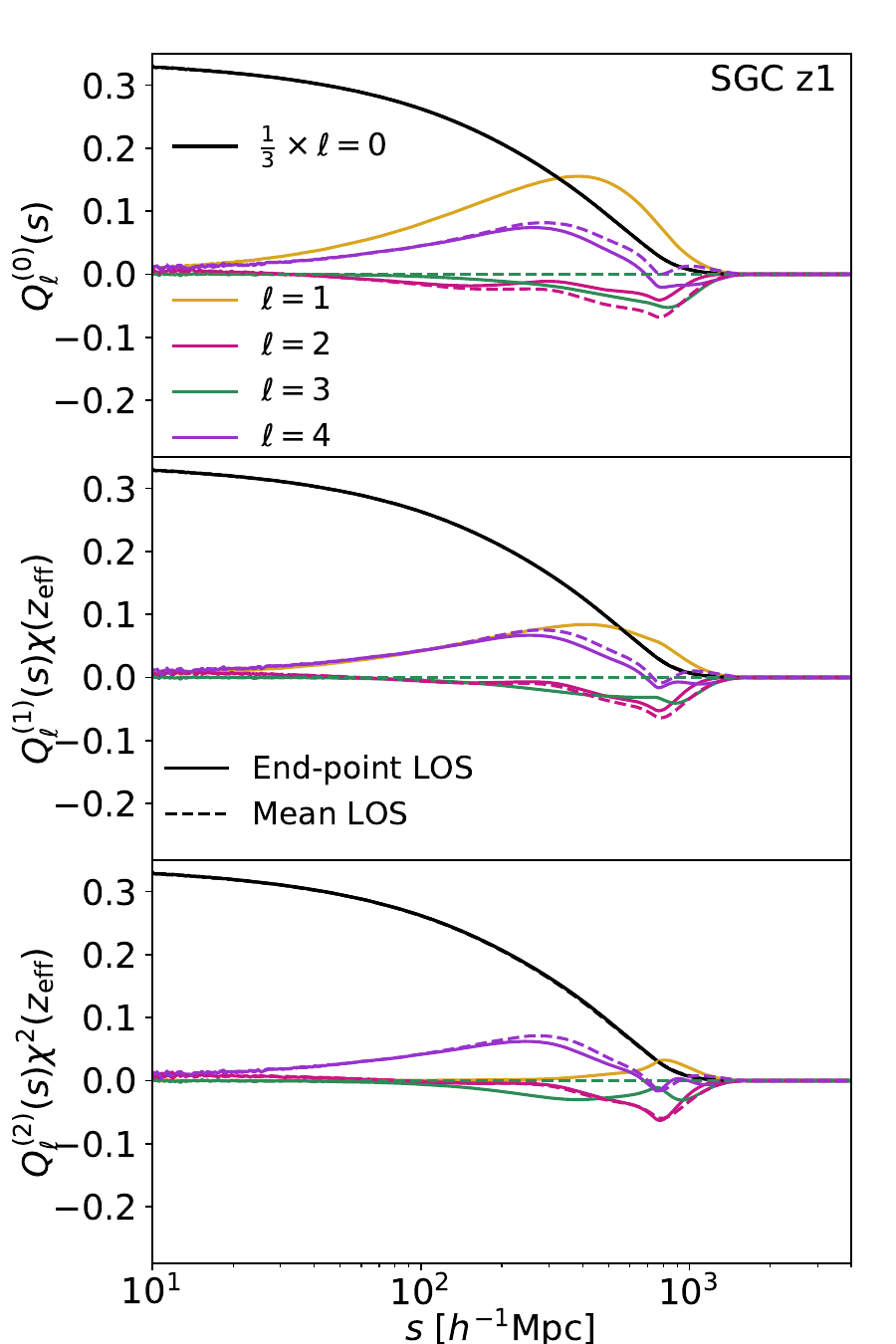}
\includegraphics[width=0.45\textwidth]{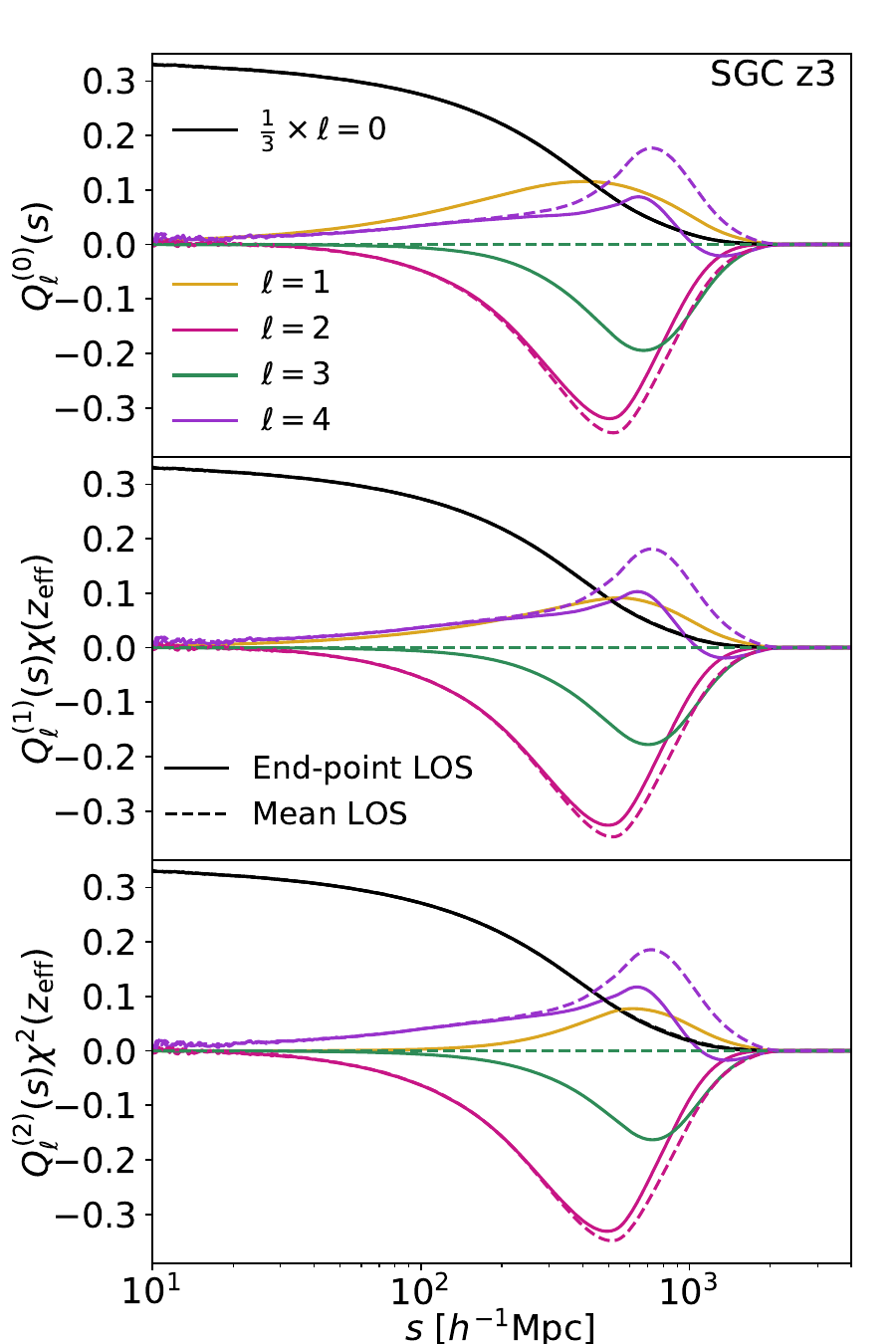}
\caption{Comparison of window function multipoles calculated by pair counting using the mean LOS definition (dashed lines) and the end-point LOS definition (solid lines) for BOSS DR12, SGC. The corresponding plots for the NGC are shown in \fig{fig:cmpLOS_NGC}. The plot on the left shows the low redshift bin of BOSS ($0.2 < z < 0.5$, $z_{\rm eff} = 0.38$), while the plot on the right shows the high redshift bin ($0.5 < z < 0.75$, $z_{\rm eff} = 0.61$). The three panels show the window functions with different orders of the wide-angle expansion, $n$, where $\chi(z_{\rm eff}=0.38)=1034.8\Mpc$ and $\chi(z_{\rm eff}=0.61)=1560.5\Mpc$ represent the co-moving distances to the effective redshift of the low (z1) and high (z3) redshift bin, respectively. The monopole term is divided by 3 to improve the visibility of the higher order multipoles. The end-point LOS definition introduces non-zero odd multipoles.}
\label{fig:cmpLOS_SGC}
\end{figure}

\subsubsection{Multipole decomposition of the window function}

Previous analyses of BOSS data (e.g.~\citep{Beutler2016:1607.03150v1,Beutler2016:1607.03149v1,Grieb,Gil2016}) have so far ignored wide-angle effects and employed the following definition of the multipoles of the window function
\begin{align}
Q_L^{(n)}(s) \equiv (2L+1) \int \mathrm{d} \Omega_s \int \mathrm{d}^3 s_1  (s_1)^{-n} W(\vb{s}_1)W(\vb{s}+\vb{s}_1)\mathcal{L}_L(\vbunit{s}\cdot \vbunit{s}_h),
\label{eq:Qn}
\end{align}
with $\vb{s}_h = (\vb{s}_1+\vb{s}_2)/2$, \ie the mean LOS. This means the LOS definition of the power spectrum estimator (based on FFTs and hence the end-point LOS) and the window function are inconsistent. The following two Sections will aim to quantify this effect. While our results will clearly show that the impact on the current BOSS RSD and BAO constraints presented in~\citep{Beutler2016:1607.03150v1,Beutler2016:1607.03149v1,Grieb,Gil2016} is negligible, future constraints on primordial non-Gaussianity or measurements of the GR induced dipole, require a consistent treatment.

We expect different LOS definitions in the window function multipoles to matter more than for the galaxy correlation function, since the survey mask does not share any of the symmetries of the underlying distribution of galaxies.
For instance there is no reason to assume odd multipoles of the window functions are much smaller then the even ones, \eg $Q_{1,3}(s)\ll Q_0(s)$, as is the case for the galaxy multipoles where $\xi_{1,3}(s)\ll \xi_0(s)$.

\fig{fig:cmpLOS_NGC} and~\ref{fig:cmpLOS_SGC} show the window function multipoles of BOSS DR12, comparing the mean LOS definition (Eq.~\ref{eq:Qn}, dashed lines) and the end-point LOS definition (Eq.~\ref{eq:Qnep}, solid lines). In these figures we removed the suppression $\chi(z_{\rm eff})^n$ of the higher order terms, where $\chi(z_{\rm eff})$ is the co-moving distance to the effective redshift of the sample. The monopole is, by definition, independent of the LOS at $n=0$, but for $\ell > 0$ the two definitions give very different results on scales larger than a few hundred $\Mpc$. While the mean LOS definition leads to even multipoles only, the end-point LOS definition results in odd multipoles, dipoles and octopoles, with similar size compared to the even multipoles.

Note that the amplitude of $\ell>0$ multipoles varies significantly between the different patches. While the low redshift bin (z1) in the South Galactic Cap (SGC) has very small higher order multipoles, indicating a close to isotropic geometry, the high redshift bin of the NGC shows the highest degree of anisotropy. We find that the window function multipoles of the high redshift bin (z3) are generally larger than in the low redshift bin (z1), which is most likely connected to the redshift distribution. This implies that the window function corrections in BOSS DR12 are generally larger for the high redshift bin, while the same is not true for the wide-angle corrections ($n>0$), since the high redshift wide-angle  corrections have an additional suppressed of $[\chi(z_1)/\chi(z_3)]^n$ compared to the low redshift bin.

\subsubsection{The convolved power in the plane-parallel limit}
\label{sec:formerwork}

\begin{figure}[t]
\centering
\includegraphics[width=1.\textwidth]{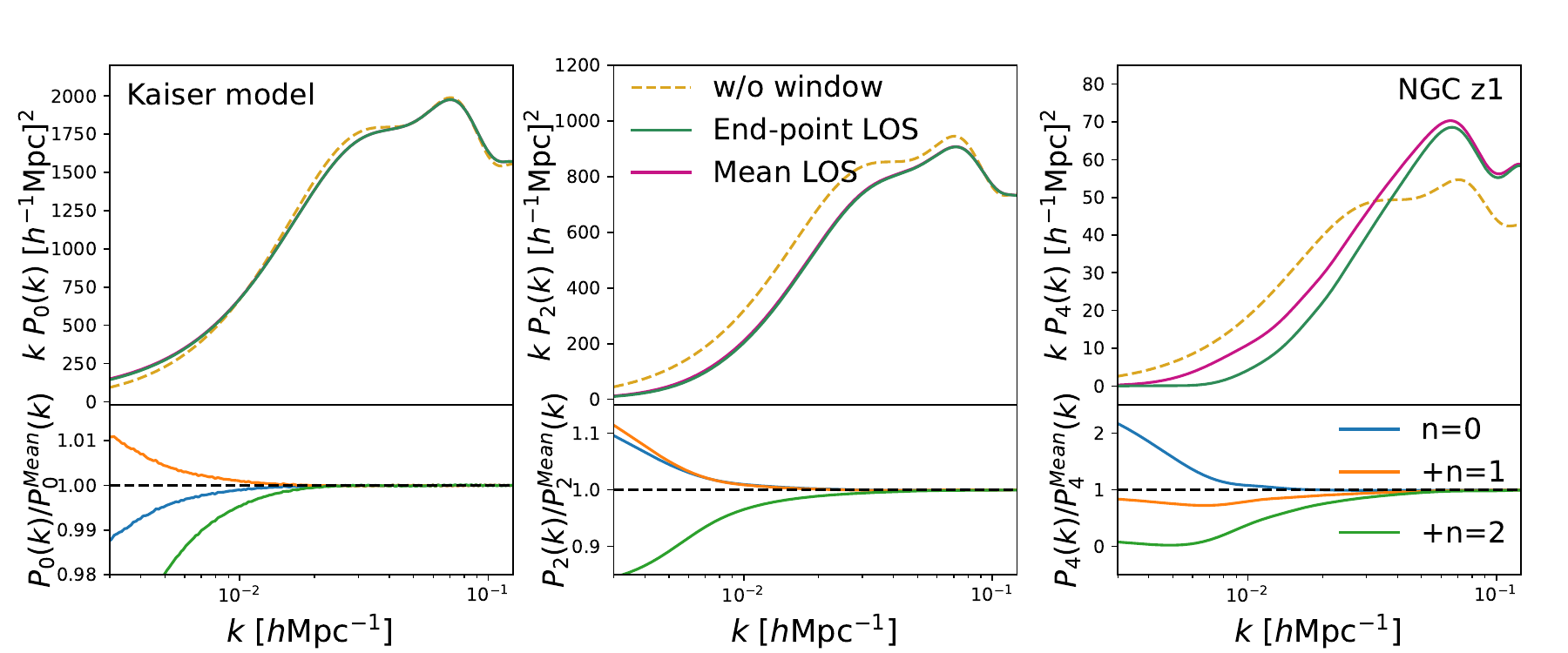}
\caption{Comparison of the power spectrum multipoles (monopole, quadrupole and hexadecapole) based on a linear Kaiser model without any window function (yellow dashed line), including the window function based on the mean LOS definition at level $n=0$ (magenta solid line) and our window function approach using the end-point LOS definition including wide-angle corrections at order $n=[0,1,2]$ (green solid line). Here we use the window function of BOSS DR12 NGC in the low redshift bin (z1). The lower panel shows the relative power spectra including each correction level $n$ in turn.}
\label{fig:old_new_cmp}
\end{figure}

In \fig{fig:cmpLOS_NGC} and~\ref{fig:cmpLOS_SGC} we have seen that different LOS definitions lead to very different window function multipoles.  
We can now take the measurements of $Q_\ell^{\rm ep}(s)$ and $Q_\ell^{\rm Mean}(s)$ from the last section and plug them into \eq{eq:convpower} to obtain the convolved power spectrum.

For illustration purposes only we assume the underlying correlation function is well represented by a linear Kaiser model, assuming a galaxy bias of $b=1.8$. This is an approximation, nonetheless the difference between the two possible choices of the LOS manifest at large scales and thus we expect linear theory to be quite accurate.

The result is shown in the three plots of \fig{fig:old_new_cmp} for the monopole (left), quadrupole (middle) and hexadecapole (right). In the top panels, the yellow dashed line shows the Kaiser model without any window function, while the magenta line shows the power spectrum with the window function calculated using the mean LOS. The green line uses the end-point LOS definition for the window function as discussed in Section~\ref{sec:wideangle}. In the lower panels, the blue line shows the relative error introduced by using different LOS definitions at order $n=0$. The reference in this case is the power spectrum model with the mean LOS, $P_L^{\rm Mean}(k)$. For the monopole the bias introduced by an inconsistent treatment of the window function leads to a $1\%$ bias in the estimated power at large scales ($k = 0.003\kMpc$), which becomes $10\%$ for the quadropole. Finally, the bottom panel in the third plot shows that comparing the measured hexadecapole to a theoretical model using the wrong definition of the LOS introduces $100\%$ inaccuracies at $k = 0.003\kMpc$. We note however, that current RSD and BAO studies use $k_{\rm min} = 0.01\kMpc$~\citep{Beutler2016:1607.03150v1,Beutler2016:1607.03149v1}, which significantly reduces the impact of these effects and makes them negligible when assuming the BOSS uncertainties. A similar analysis for Fourier space clustering wedges~\cite{Grieb} can be found in Appendix~\ref{sec:wedges}.

Comparing the set of lines in the bottom panels, we can anticipate that the error introduced by using inconsistent LOS definition for the power spectrum multipoles and the window functions, is of the same order as the wide-angle correction terms discussed in the next section. In the case of BOSS, the errors shown in \fig{fig:old_new_cmp} are well below the measurement uncertainties and hence we do not expect the findings of this section to impact those past analyses. 

\subsection{The wide-angle power spectra}

\eq{eq:convpower} shows how different $\ell$-multipoles of the underlying correlation function contribute to the mean of the estimated $A$-multipole of the power spectrum through couplings with the $L$-multipoles of the window function.
The structure of these couplings is dictated by the $3j$ symbols, similarly to what happens in the CMB case~\cite{Hiv02}, and, most importantly, it is independent of the order $n$ at which the wide-angle terms are computed. 
While we discuss the explicit expressions for the odd multipoles in Section~\ref{sec:P1} and~\ref{sec:P3}, the corresponding expressions for the even multipoles are given in Appendix~\ref{App:even}.

\subsubsection{Even multipoles}

\begin{figure}[t]
\centering
\includegraphics[width=0.32\textwidth]{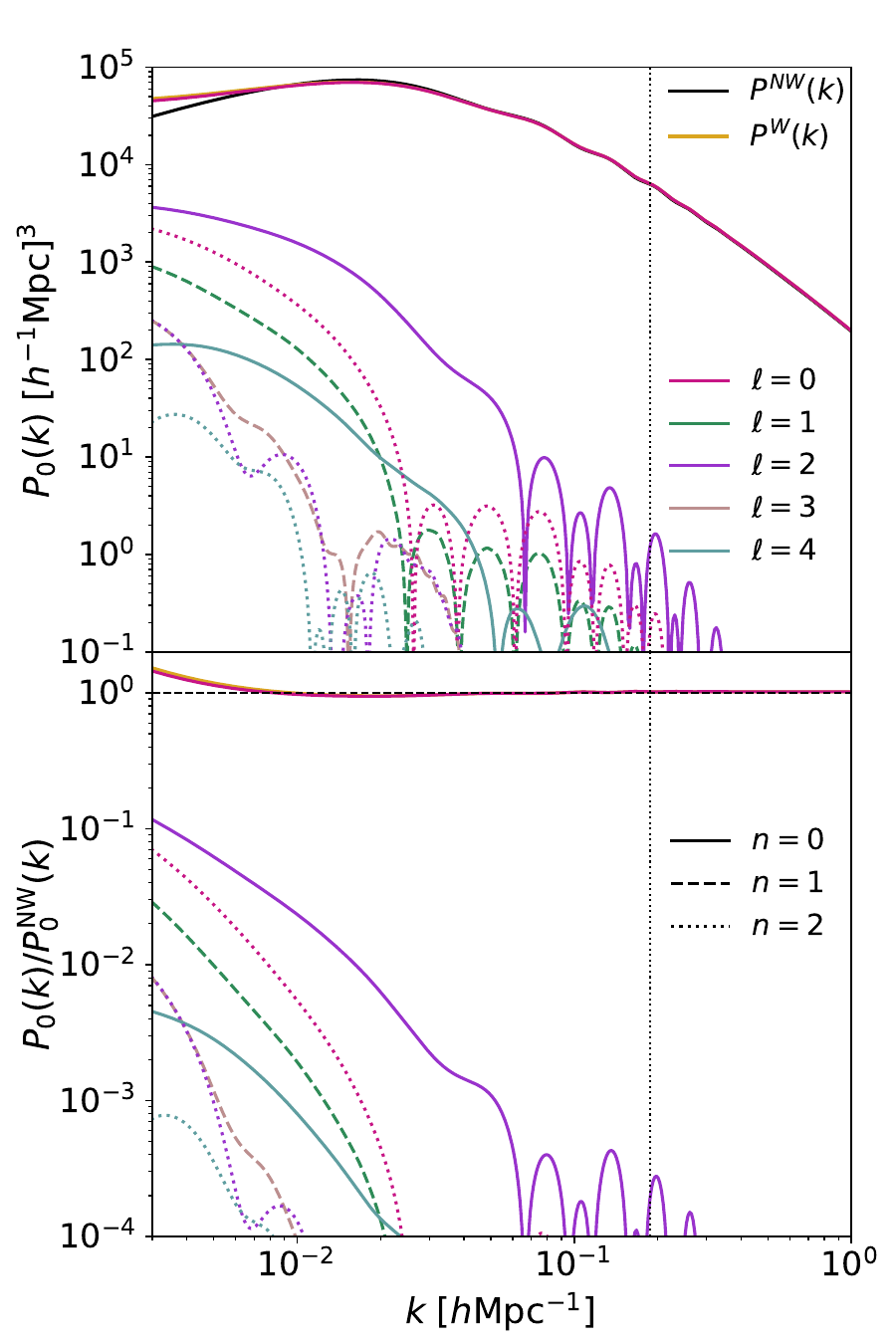}
\includegraphics[width=0.32\textwidth]{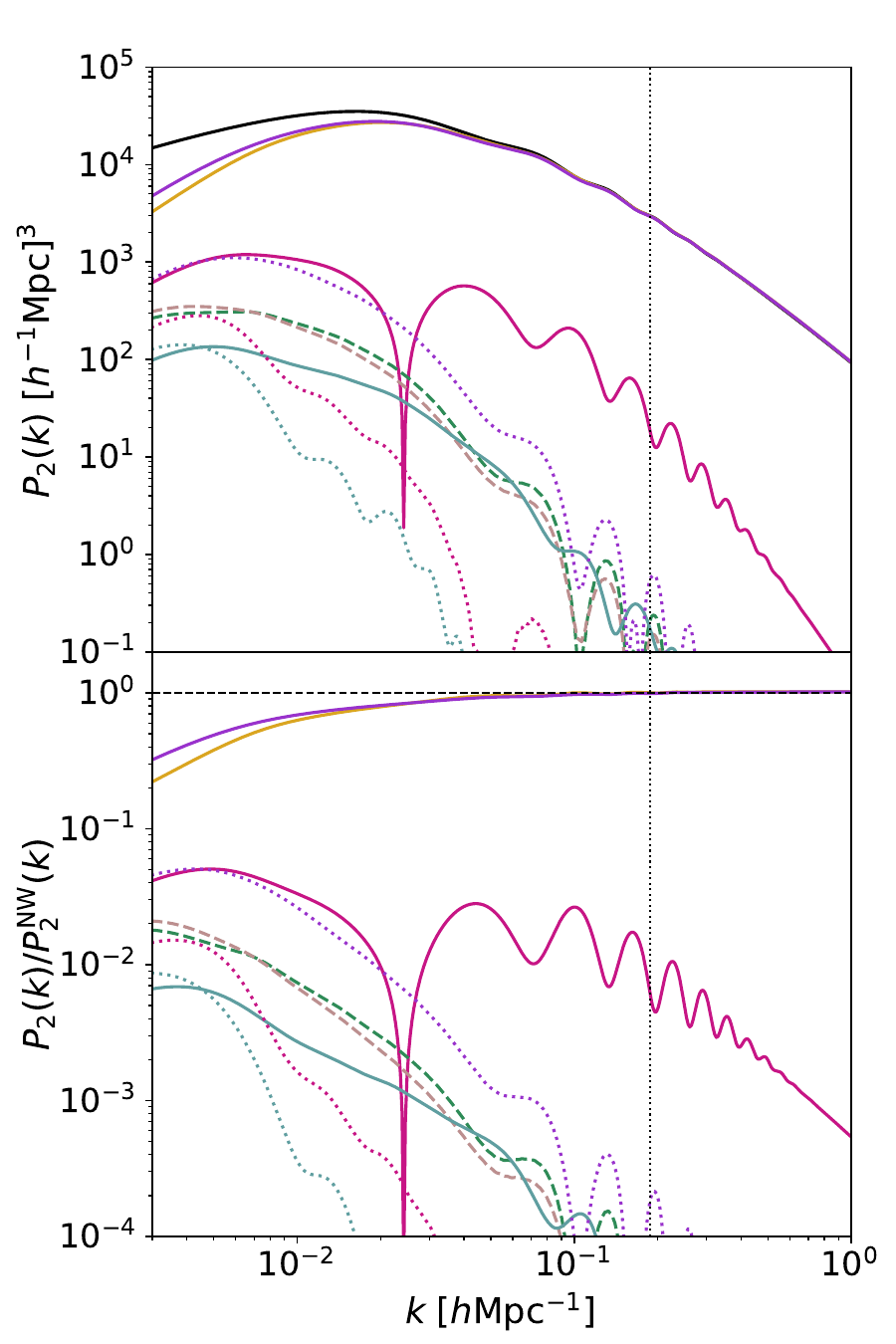}
\includegraphics[width=0.32\textwidth]{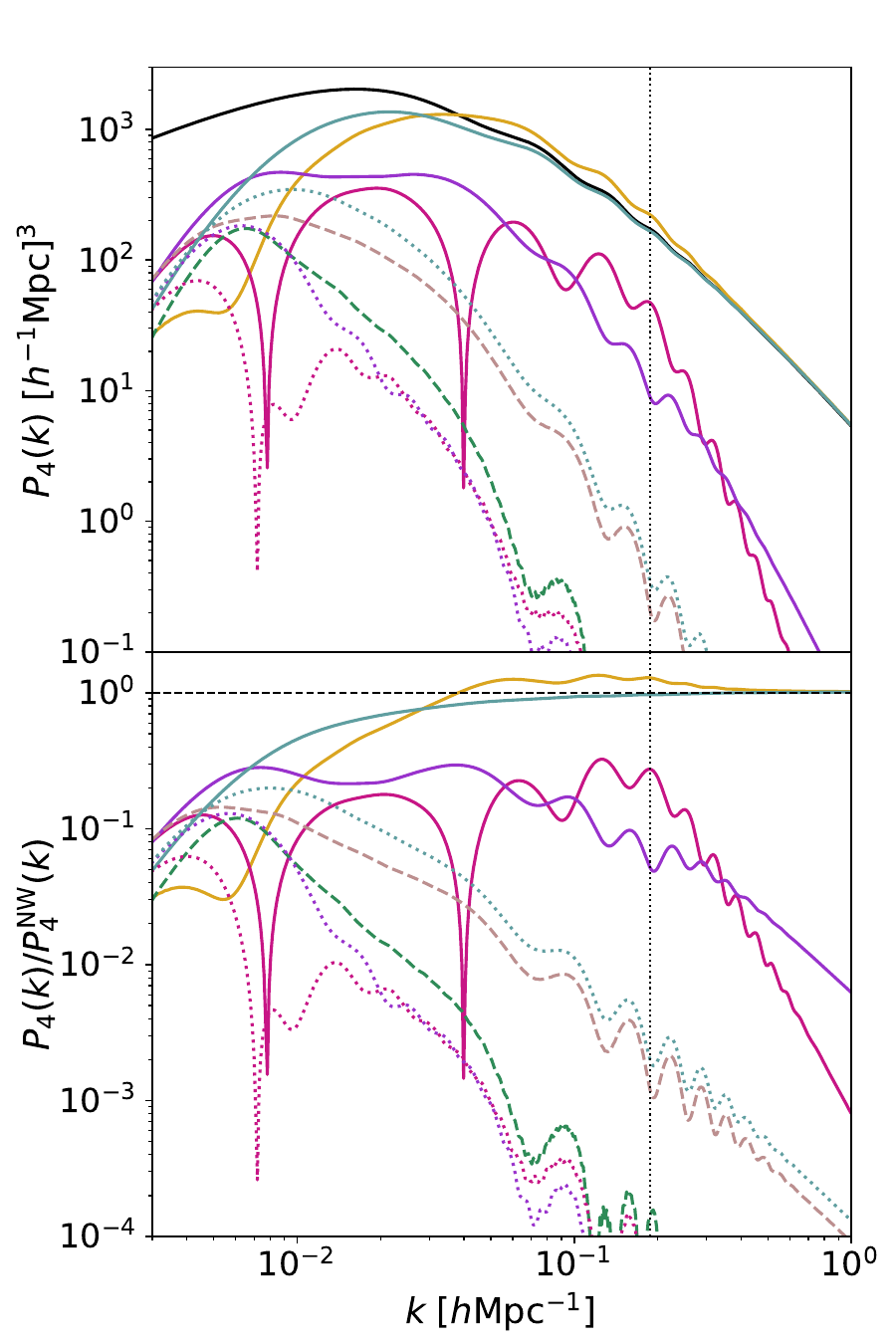}
\caption{Comparison of the terms sourced by the interactions between the wide-angle effect and the window function for the even multipoles of the low redshift bin (z1) of BOSS DR12, NGC. The odd multipoles are shown in \fig{fig:individual_odd_NGC_z1}. All terms are calculated within the linear Kaiser model. The black solid line shows the power spectrum multipole without any window function effect, while the yellow solid line shows the same model including all wide-angle corrections of order $n < 3$. The lower panel shows the relative contributions of the different terms. Note that we always plot the absolute value $|P_{\ell}(k)|$. The line style indicates the order ($n$) of the wide-angle correction with $n=2$ (dotted lines) generally being above the $n=1$ terms (dashed lines), while the $n=0$ terms (solid lines) dominate. The vertical dotted line indicates the Nyquist frequency of the window function at $k_{\rm ny} = 0.19\kMpc$, below which the measured window functions are smoothed out (see Appendix~\ref{sec:FFTwindow}).}
\label{fig:individual_even_NGC_z1}
\end{figure}

For the monopole, $A=0$, it is easy to see that in \eq{eq:convpower} the only non-vanishing contributions arise when $\ell=L$,
\begin{align}
	\convolved{P}_0(k) &= \sum_{\ell}\tj{\ell}{\ell}{0}{0}{0}{0}^2\int \mathrm{d}s\, j_0 (ks) \sum_n s^{n+2}\xi_\ell^{(n)}(s)  Q_\ell^{(n)}(s) \\ 
	&\equiv\sum_n \int ds\, j_{0}(ks)s^{n+2} \convolved{\xi}_{0}^{(n)}(s)\,.
\end{align}
These monopole correction terms are shown in \fig{fig:individual_even_NGC_z1}, left panel. The black line shows the power spectrum without any window function $P_0^{NW}(k)$, the yellow line shows the convolved power $P^W(k)$, and the other lines show the individual contributions from the different multipoles with the linestyle indicating the level $n$ in the wide-angle correction. As in Section~\ref{sec:formerwork} we use linear theory with galaxy bias $b=1.8$.

We find that the measured monopole is dominated by the underlying $\ell=0$ power spectrum, simply due to the fact that the $Q_0^{(0)}(s)$ is much larger than the other multipoles of the window function. 
The second largest term is the quadrupole at $n=0$, which contributes at the $10\%$ level at large scales.
At first order in the wide-angle corrections the monopole receives non-negligible contributions, around 2\% of the total power, from the dipole and the octopole,
\begin{align}
	\convolved{\xi}_0^{(1)}(s) = \frac{1}{3}\xi_1^{(1)}(s)Q_1^{(1)}(s)+\frac{1}{7}\xi_3^{(1)}(s)Q_3^{(1)}(s) + \cdots.
\end{align}
We also notice that in a variety of cases the wide-angle contributions can be comparable or larger than the $n=0$ terms, for instance the $n=2$ contribution from $\ell=0$, red dotted line in \fig{fig:individual_even_NGC_z1}, is much bigger than the hexadecapole at $n=0$, in green. The relative contributions of the different terms are shown in the lower panel of \fig{fig:individual_even_NGC_z1}.

Accounting for the $n=1$ contributions partially cancels the bias coming from an inconsistent treatment of the window function discussed in Section~\ref{sec:formerwork}, as one can see by comparing the orange and blue lines in \fig{fig:old_new_cmp}. Including $n=2$ terms results in a $3\%$ systematic underestimate of the monopole power spectrum at large scales.

Similar conclusions can be drawn for the convolved quadrupole and hexadecapole.
The quadrupole, $A=2$, receives the largest off diagonal, \ie $\ell\ne A$, contribution from $\ell=0$, with both $n=0$ and $n=2$ comparable in size at low $k$ (see the middle panel in \fig{fig:individual_even_NGC_z1}). 
The dipole and the octopole terms make up $1\%$ of the total power and are much larger then the $\ell=4$ piece.
As expected wide-angle effects are more important for higher order multipoles, with the lower panel in the middle plot of \fig{fig:individual_even_NGC_z1} showing that neglecting $n=1,2$ in computing the mean of the FFT-estimator for the quadrupole leads to $20\%$ biases at large scales.

Finally, the hexadecapole is the most severely affected by the survey geometry and the wide-angle terms.
As shown in the right plot of \fig{fig:individual_even_NGC_z1}, several multipoles give rise to sizable effects at large scales, at any order in $n$.
At $k\le 10^{-2} \kMpc$, the power in the hexadecapole is spread out among almost $10$ different terms.
Neglecting wide-angle terms, and their correlation with the window function multipoles, leads to more than a $100\%$ bias for the hexadecapole at $k=0.003\kMpc$ in the case of BOSS.

\begin{figure}[t]
\centering
\includegraphics[width=0.485\textwidth]{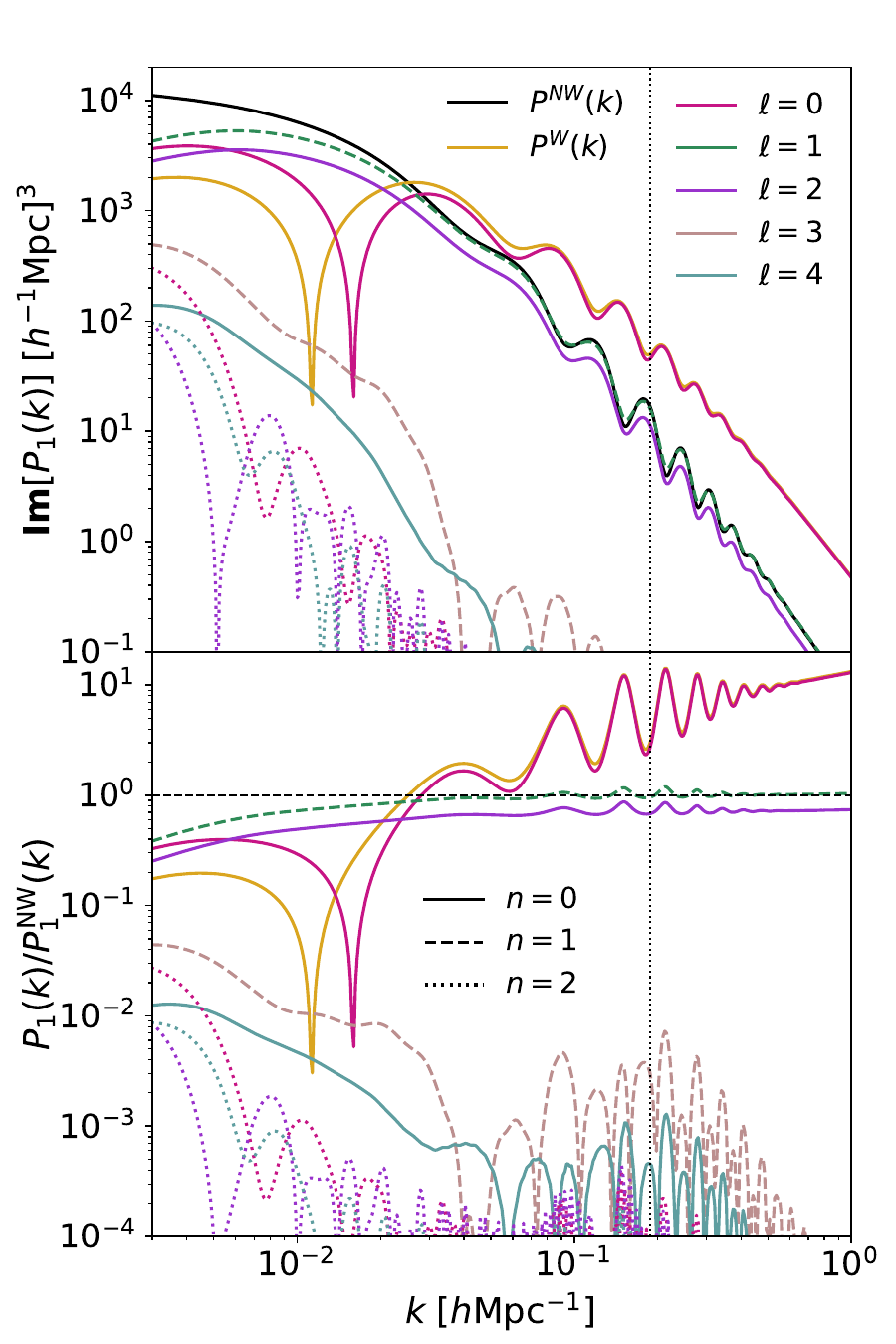}
\includegraphics[width=0.485\textwidth]{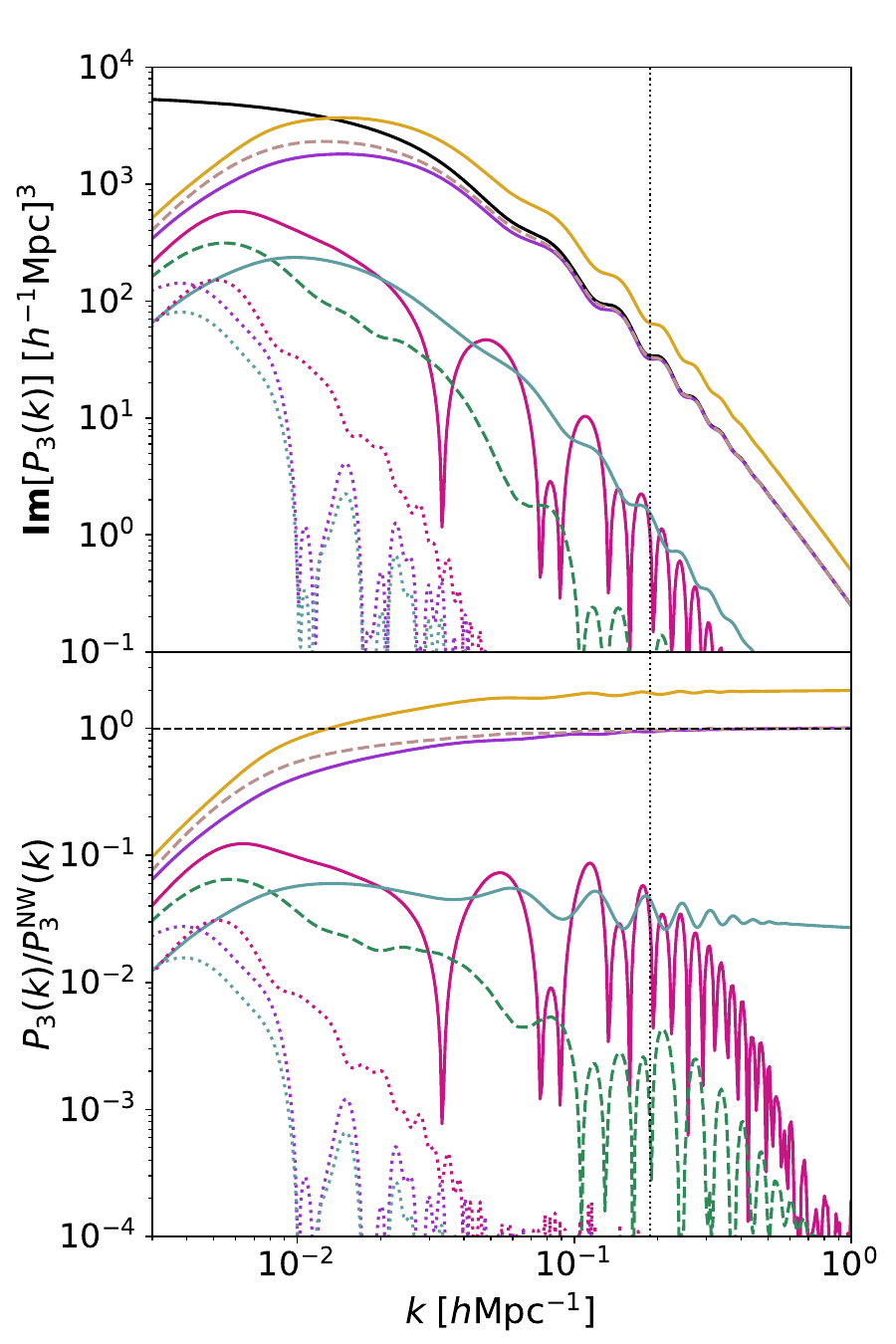}
\caption{Comparison of the terms sourced by the interactions between the wide-angle effect and the window function for the odd multipoles of the low redshift bin (z1) of BOSS DR12, NGC. The even multipoles are shown in \fig{fig:individual_even_NGC_z1}. All terms are calculated within the linear Kaiser model. The black solid line shows the power spectrum multipole without any window function effect (see Eq.~\ref{eq:order1term0} and~\ref{eq:order1term1}), while the yellow solid line shows the same model including all wide-angle corrections of order $n < 3$. The lower panel shows the relative contributions of the different terms. Note that we always plot the absolute value $|P_{\ell}(k)|$. The line style indicates the order ($n$) of the wide-angle correction, with the $n=0$ contributions dominating on most scales. Different to the case of the even multipoles the $n=1$ terms (dashed lines), are usually far above the $n=2$ terms (dotted lines). The vertical dotted line indicates the Nyquist frequency of the window function at $k_{\rm ny} = 0.19\kMpc$, below which the measured window functions are smoothed out (see Appendix~\ref{sec:FFTwindow}).}
\label{fig:individual_odd_NGC_z1}
\end{figure}

\subsubsection{The power spectrum dipole, $P_1(k)$}
\label{sec:P1}

The presence of odd multipoles is a consequence of the asymmetric choice of the LOS in the FFT-estimator. These geometric terms do not carry any extra cosmological information, but they can be used as a consistency check of the data analysis and to learn about the properties of the window function.
Geometric dipoles also represent a contamination in the search for relativistic effects and should therefore be modeled properly.

The convolved power spectrum dipole reads
\begin{align}
	\convolved{P}_1(k) =& -3i \sum_{\ell,\, L,n}\tj{\ell}{L}{1}{0}{0}{0}^2\int \mathrm{d}s\,s^{n+2} j_1 (ks) \xi_\ell^{(n)}(s)  Q_L^{(n)}(s) \notag \\
    \equiv& -3 i \sum_n \int \mathrm{d}s\,s^{n+2} j_1 (ks) \,\tilde{\xi}_1^{(n)}(s)
\end{align}
from which one sees that, although the dipole is intrinsically a $n=1$ term as shown in \eq{eq:order1term1}, the presence of the angular mask generates a dipole even in the plane-parallel limit. 
We define the convolved correlation function dipole as
\begin{align}
	\begin{split}
		\convolved{\xi}_1^{(0)}(s) = \xi^{(0)}_0(s)Q_1^{(0)}(s)&+\xi^{(0)}_2(s)\left[\frac{2}{5}Q_1^{(0)}(s)+\frac{9}{35}Q_3^{(0)}(s)\right]\\
		&+ \xi^{(0)}_4(s)\left[\frac{4}{21}Q_3^{(0)}(s)+\frac{5}{33}Q_5^{(0)}(s)\right]\\
		&+\cdots
	\end{split}
    \label{eq:xi01}
\end{align}
This equation shows how the measured dipole receives a contribution from the coupling of the monopole of the underlying galaxy distribution with the dipole of the survey mask, which we have measured to be quite large in \fig{fig:cmpLOS_SGC}. Since the intrinsic dipole is proportional to $\xi_2^{(0)}(s)$ (see Eq.~\ref{eq:order1term1}) the first term in the square bracket of the above equation will also generate power on scales where $Q_1^{(0)}(s)\ge Q_0^{(1)}(s)$. Similar, but sub-dominant terms, can be written for $n=2$. At leading order in wide-angle correction we get the following structure of terms
\begin{align}
	\begin{split}
		\convolved{\xi}_1^{(1)}(s) &= \xi_1^{(1)}(s)\left[Q_0^{(1)}(s) +\frac{2}{5} Q_2^{(1)}(s)\right]\\
		&+\xi_3^{(1)}(s)\left[\frac{9}{35}Q_2^{(1)}(s) +\frac{4}{21}Q_4^{(1)}(s)\right]\\
		&+\cdots
	\end{split}
\end{align}
All dipole related terms are shown in the left plot of \fig{fig:individual_odd_NGC_z1}. As anticipated, the presence of large dipoles in the survey geometry makes the $\ell=0,2$ contributions at $n=0$ (red and purple lines) larger than, or comparable to, the intrinsic one, shown with the dashed green line.
This appears to be the case for all redshifts and patches of the sky, NGC and SGC, we analyzed.
Our results show how a proper knowledge of the window function will be crucial in any attempted detection of general relativistic dipoles at small and large scales.

\subsubsection{The power spectrum octopole, $P_3(k)$}
\label{sec:P3}

The story pretty much repeats itself for the measured octopole,
\begin{align}
	\convolved{P}_3(k) =& -7i \sum_{\ell,\, L,n}\tj{\ell}{L}{3}{0}{0}{0}^2\int \mathrm{d}s\,s^{n+2} j_3 (ks) \xi_\ell^{(n)}(s)  Q_L^{(n)}(s) \notag \\
  \equiv& -7 i \sum_n \int \mathrm{d}s\,s^{n+2} j_3 (ks) \,\tilde{\xi}_3^{(n)}(s),
\end{align}
which has the following structure at $n=0$
\begin{align}
	\begin{split}
		\convolved{\xi}_3^{(0)}(s) = \xi^{(0)}_0(s)Q_3^{(0)}(s)&+\xi^{(0)}_2(s)\left[\frac{3}{5}Q_1^{(0)}(s)+\frac{4}{15}Q_3^{(0)}(s)+\frac{10}{33}Q_5^{(0)}(s)\right] \\ 
&+\xi^{(0)}_4(s)\left[\frac{4}{9}Q_1^{(0)}(s)+\frac{2}{11}Q_3^{(0)}(s)+\frac{20}{143}Q_5^{(0)}(s)\right] \\
        & + \cdots
	\end{split}
\end{align}
and at $n=1$
\begin{align}
	\begin{split}
		\convolved{\xi}_3^{(1)}(s) &= \xi_1^{(1)}(s) \left[\frac{3}{5} Q_2^{(1)}(s) +\frac{4}{9} Q_4^{(1)}(s)\right]\\ 
        & +\xi_3^{(1)}(s)\left[Q_0^{(1)}(s)+\frac{4}{15}Q_2^{(1)}(s) +\frac{2}{11}Q_4^{(1)}(s)\right]\\
        & + \cdots
	\end{split}
    \label{eq:xi31}
\end{align}
Similarly to the dipole in the previous section, $n=0$ terms are produced by the coupling of the underlying theory with the odd multipoles of the survey mask. The right plot of \fig{fig:individual_odd_NGC_z1} shows the wide-angle contributions to the octopole, which are dominated by the first term in the squared bracket of \eq{eq:xi31} on most scales.

\subsection{The integral constraint}

\begin{figure}[t]
\centering
\includegraphics[width=0.45\textwidth]{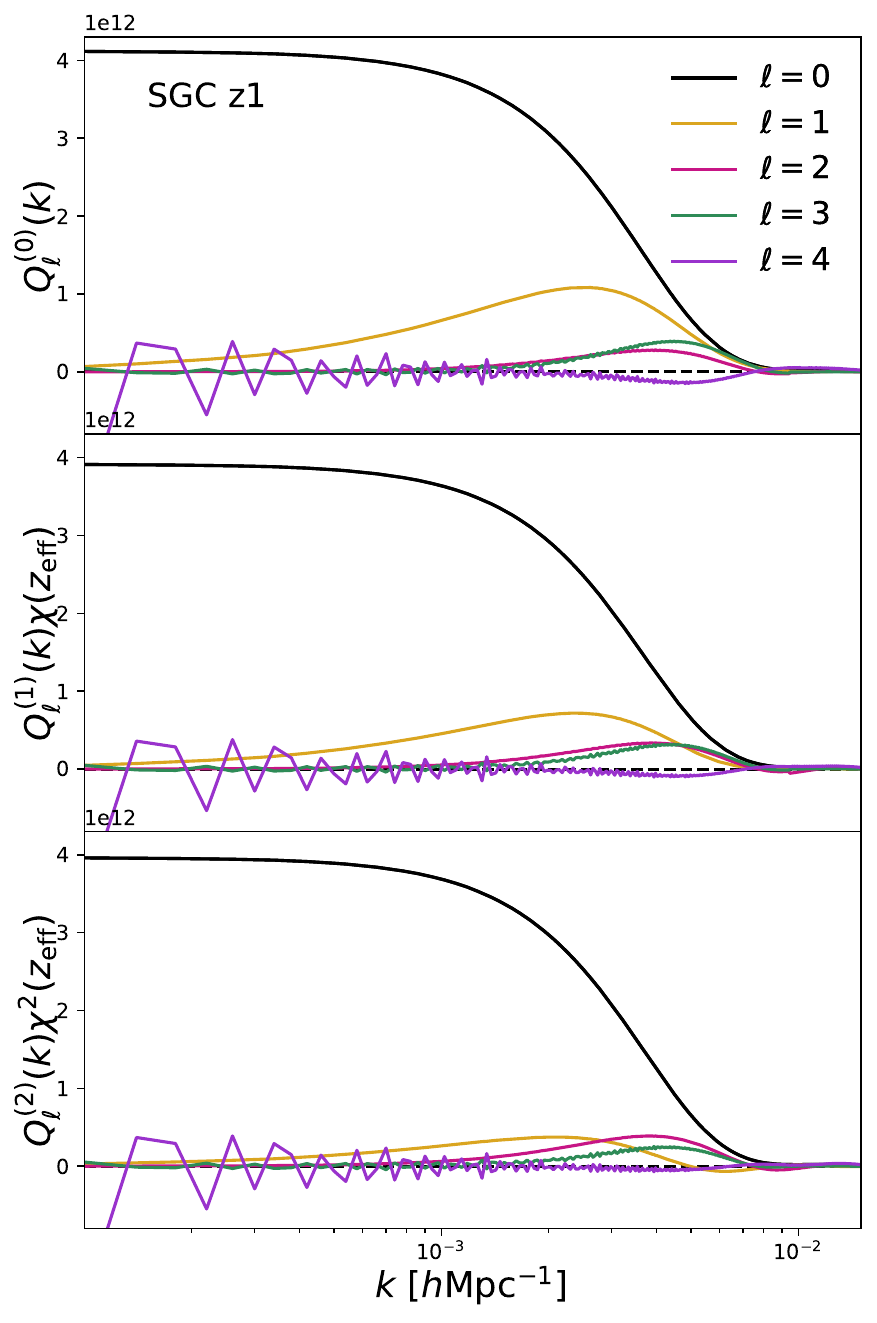}
\includegraphics[width=0.45\textwidth]{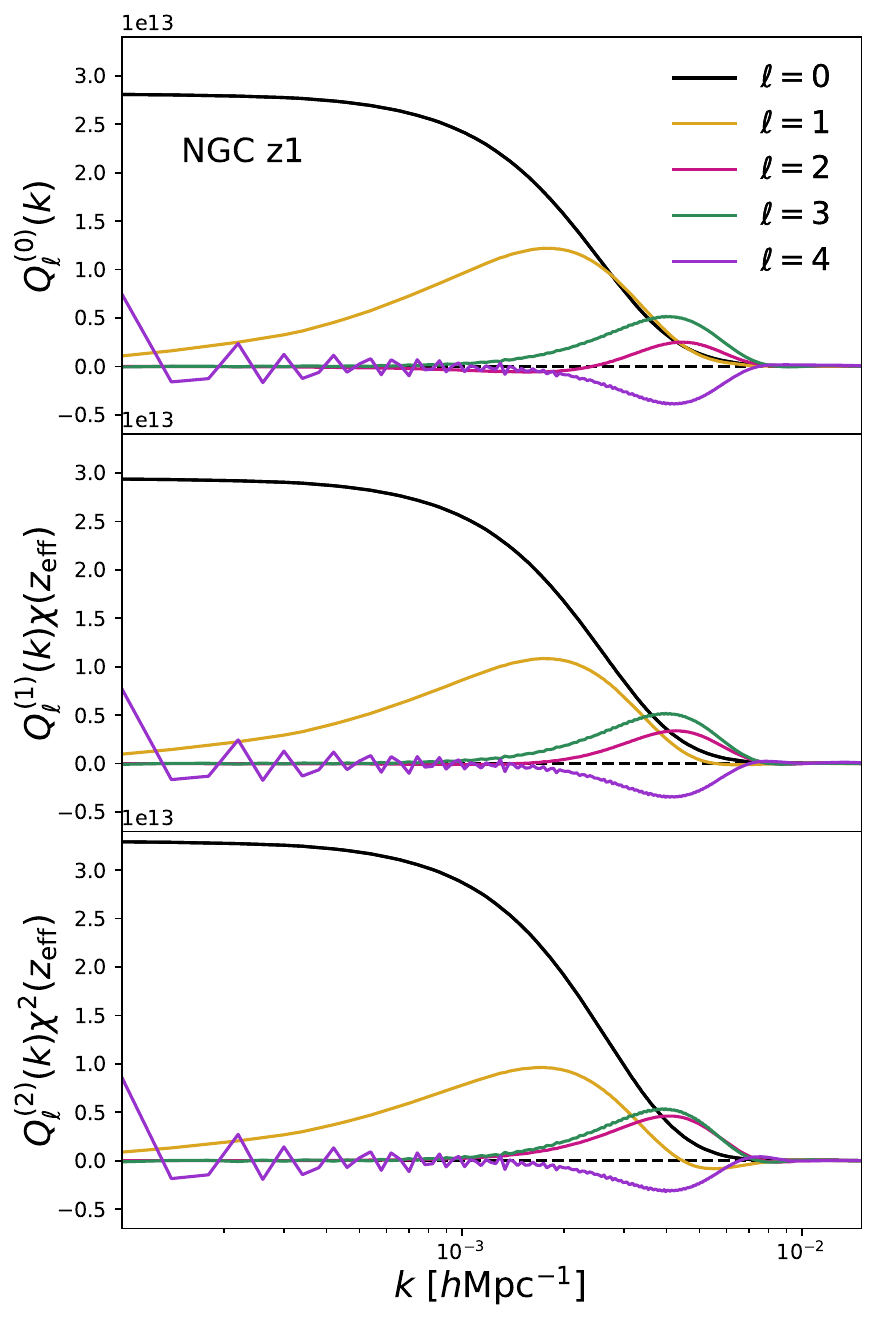}
\caption{Window function multipoles in Fourier-space for the low redshift bin ($0.2 < z < 0.5$, $z_{\rm eff} = 0.38$) of both galactic caps of BOSS DR12. The first panel shows the window function terms at order $n=0$, while the middle and lower panel show the wide-angle corrections at order $n=1$ and $n=2$, respectively. The noise at small wavenumber $k$ is caused by the lack of modes at those scales which is enhanced for the higher order multipoles, which give more weight to a smaller number of modes.}
\label{fig:kwindow}
\end{figure}

\begin{figure}[t]
\centering
\includegraphics[width=0.45\textwidth]{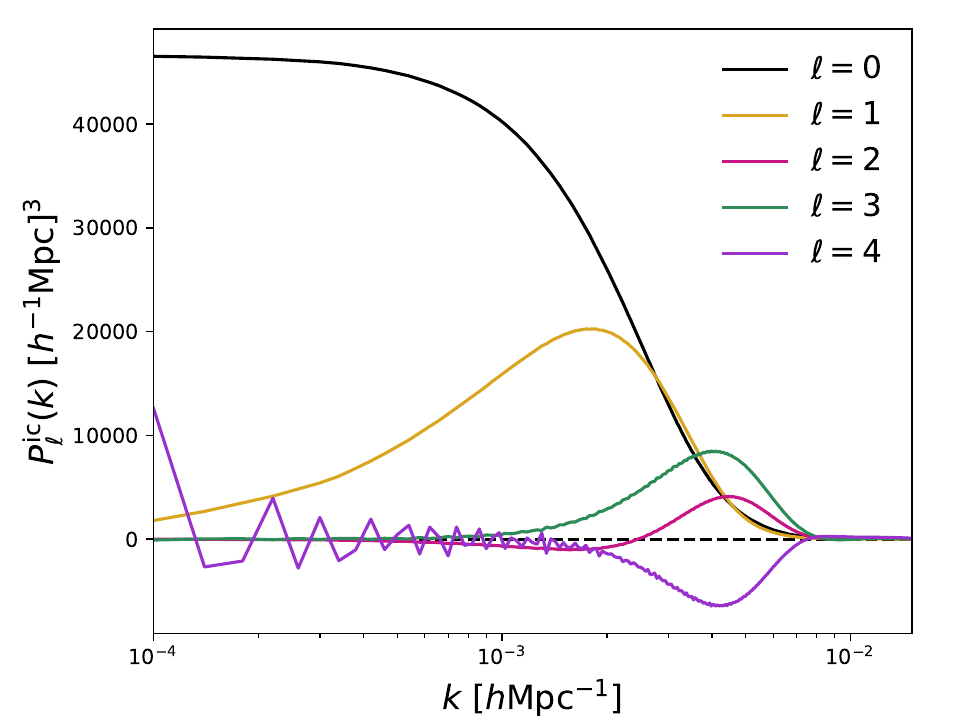}
\includegraphics[width=0.45\textwidth]{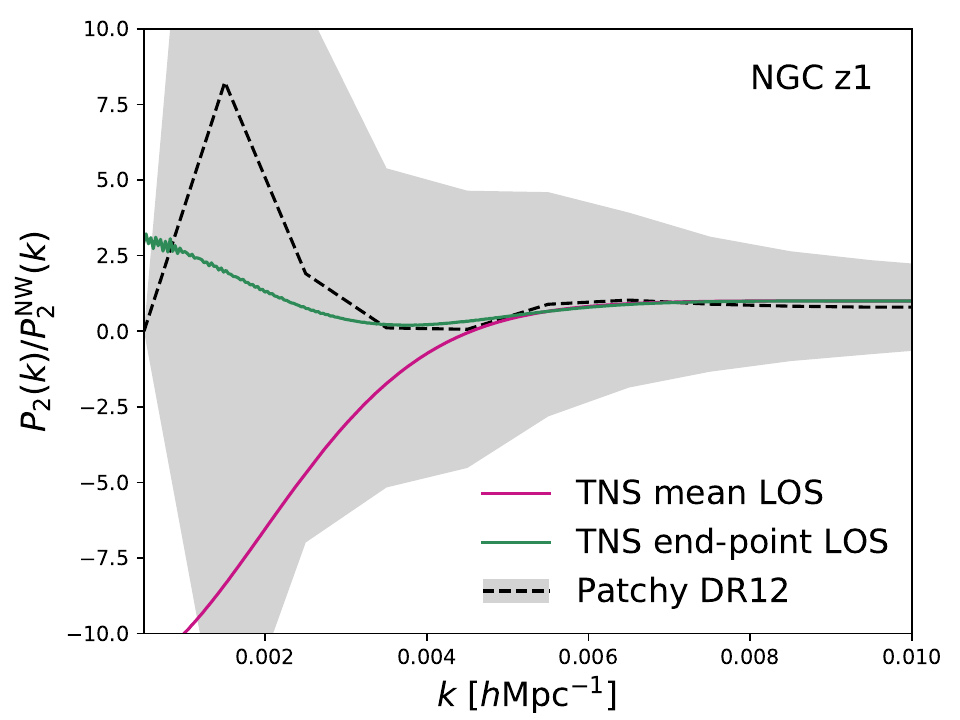}
\caption{The plot on the left shows the integral constraint contributions for the different multipoles. These terms are scaled versions of the window function multipoles shown in \fig{fig:kwindow}. The noise in the higher order terms is caused by the limited number of modes available and this effect can be reduced by increasing the box size of the FFT when calculating the window functions (see the discussion in Appendix~\ref{sec:FFTwindow}). The plot on the right shows the convolved power spectrum quadrupole relative to the unconvolved power spectrum. The magenta line uses the integral constraint correction based on the mean LOS, while the green line uses the end-point LOS definition. For comparison, the black dashed line shows the measured mean power spectrum quadrupole of the Multidark Patchy mock catalogs.}
\label{fig:ic}
\end{figure}

When defining the galaxy over-density field as input for the power spectrum estimate, we need to specify the mean galaxy density of the Universe. This mean density is usually set to the mean density of the survey, which enforces the measured power spectrum to have zero power at the scale of the survey. This procedure is known as the integral constraint. To avoid any bias in the cosmological parameters we have to enforce the same constraint on the power spectrum model. The impact of the integral constraint is in principle limited to the $k=0$ mode, however the window function correlates adjacent modes and hence the integral constraint leaks to larger $k$. 

The integral constraint is given by 
\begin{equation}
P^{\rm ic}_{\ell}(k) = \convolved{P}_{0}(0)Q^{(0)}_{\ell}(k),
\label{eq:ic}
\end{equation}
where the normalization of the k-space window is given by
\begin{equation}
Q^{(0)}_{0}(k\rightarrow 0) = 1.
\end{equation}
\eq{eq:ic} holds at any order in the wide-angle correction. We can now obtain our power spectrum model as $\convolved{P}_{\ell}(k) = \convolved{P}_{\ell}(k) - P^{\rm ic}_{\ell}(k)$. 

To calculate the integral constraint we need the Fourier-space window function, which we can obtain by Hankel transforming the configuration-space window functions. In the case of the end-point LOS definition we can also directly estimate $Q^{(n)}_{L}(k)$ in Fourier-space, just as we estimate the power spectrum. More details can be found in Appendix~\ref{sec:FFTwindow}. 

\fig{fig:kwindow} shows the window function multipoles of the low redshift bin of BOSS, NGC. The noise in some of the multipoles is caused by the lack of modes on large scales due to the finite volume, which is a more prominent effect for the higher order multipoles (see the discussion in Appendix~\ref{sec:FFTwindow}). \fig{fig:kwindow} clearly shows that the integral constraint effect reaches zero well before $k = 0.01\kMpc$, which is well outside the fitting range of most RSD or BAO based cosmological parameter analysis (e.g.~\citep{Beutler2016:1607.03150v1} imposes $k > k_{\rm min} =0.01\kMpc$). Nevertheless, it is important for low-k observables like the scale-dependent bias induced by primordial non-Gaussianity.

\fig{fig:ic} (left) shows the individual integral constraint terms for the different multipoles, which are scaled versions of the window functions shown in \fig{fig:kwindow}. The right plot of \fig{fig:ic} shows the power spectrum quadrupole relative to the quadrupole without any window function correction. The black dashed line shows the mean of $2048$ Multidark Patchy mock catalogs and the gray shaded area shows the variance between those mock realizations. Given that the integral constraint correction is proportional to the window function, it again depends on the choice of the line-of-sight. In \fig{fig:ic} we show the power spectrum using the mean and end-point LOS definition (magenta and green line, respectively). The Patchy power spectra have been measured assuming an end-point LOS, and hence the green model reflects the correct treatment of the integral constraint correction.  
The effect of the integral constraint is well within the cosmic variance of the measurements, indicated by the gray band, but the difference between the two choices of the LOS, \ie between the green and magenta lines in \fig{fig:ic}, would matter in any scenario where cross-correlation of two different samples will partially cancel cosmic variance.

\subsection{Analysis of BOSS DR12}
\label{sec:analysis}

\begin{figure}[t]
\centering
\includegraphics[width=0.65\textwidth]{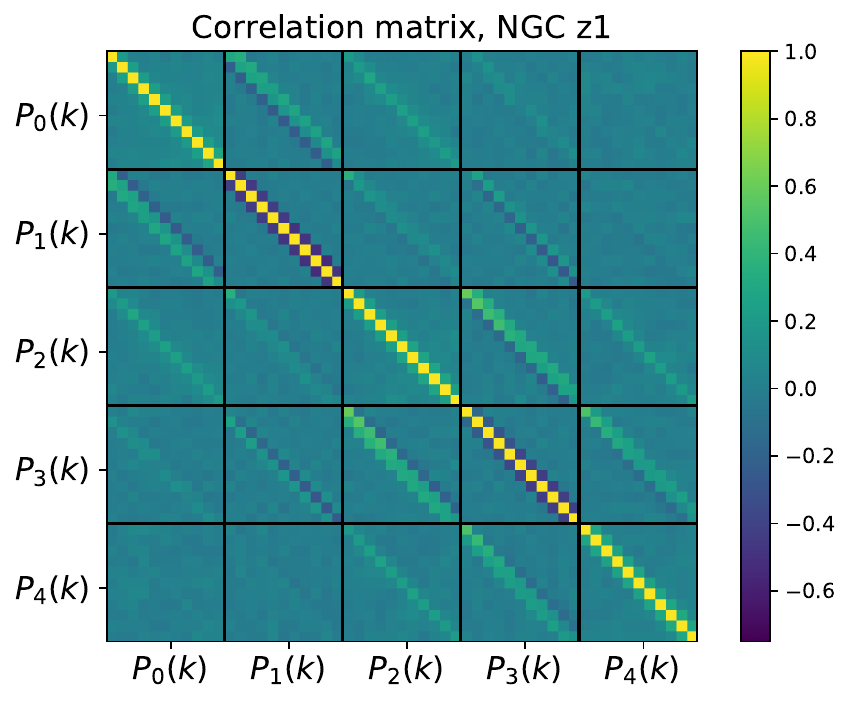}
\caption{The correlation matrix for the 5 power spectrum multipoles of the low redshift bin (z1) of BOSS NGC for the $11$ bins between $k_{\rm min} = 0.008\kMpc$ and $k_{\rm max} = 0.096\kMpc$ with $\Delta k = 0.008\kMpc$. This matrix has been derived from $2048$ Multidark Patchy mock catalogs. The color bar on the right indicates the level of correlation between bins. As expected, the dipole is strongly correlated with the monopole, which through the window function dominates the dipole (see \fig{fig:individual_odd_NGC_z1}). The octopole is correlated with the quadrupole and hexadecapole through the wide-angle sourced octopole (see Eq.~\ref{eq:order1term1}).}
\label{fig:cov_NGC_z1}
\end{figure}

\begin{figure}[t]
\centering
\includegraphics[width=0.325\textwidth]{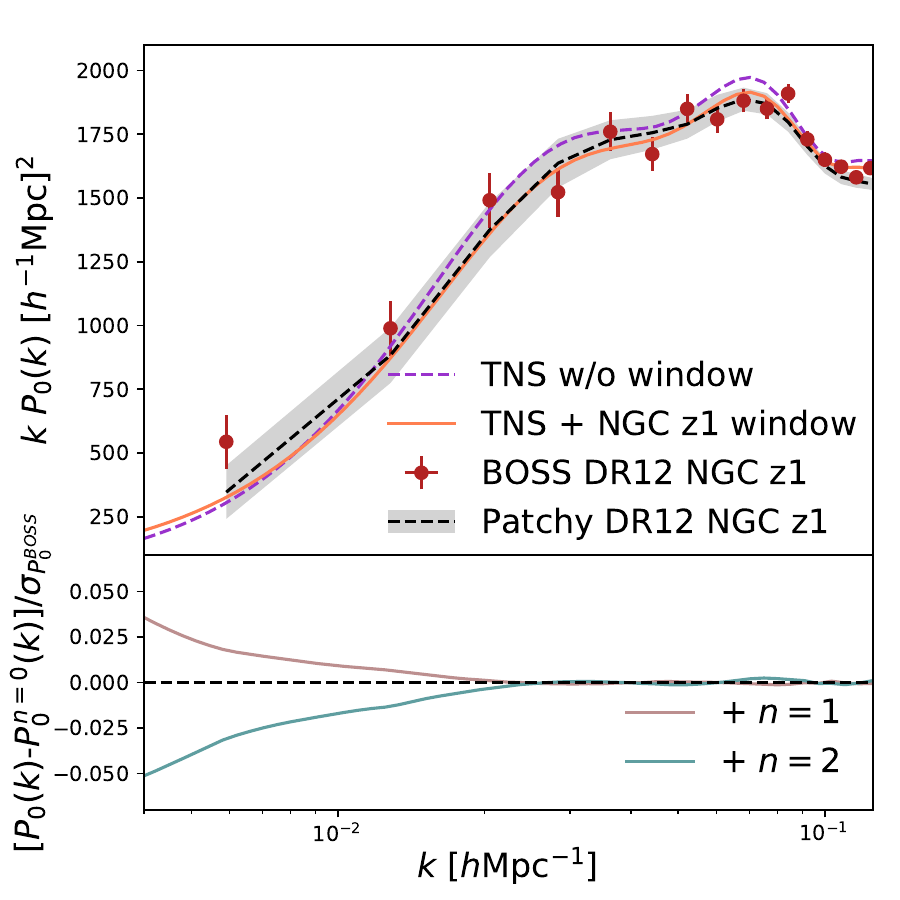}
\includegraphics[width=0.325\textwidth]{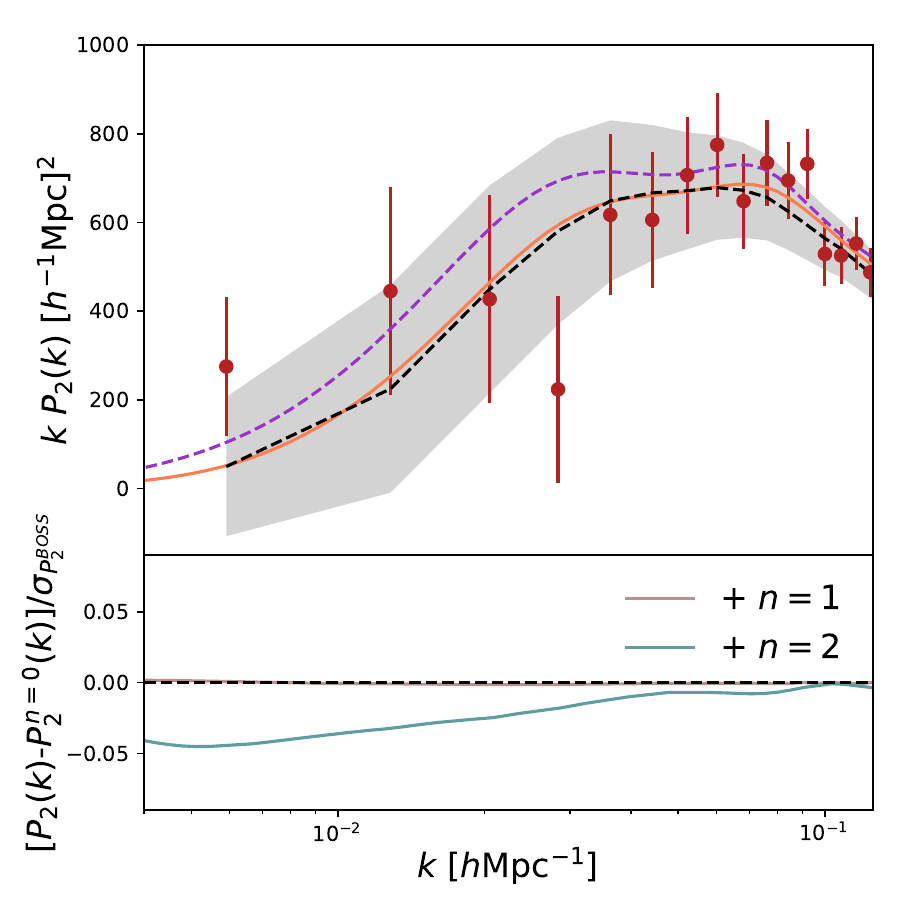}
\includegraphics[width=0.325\textwidth]{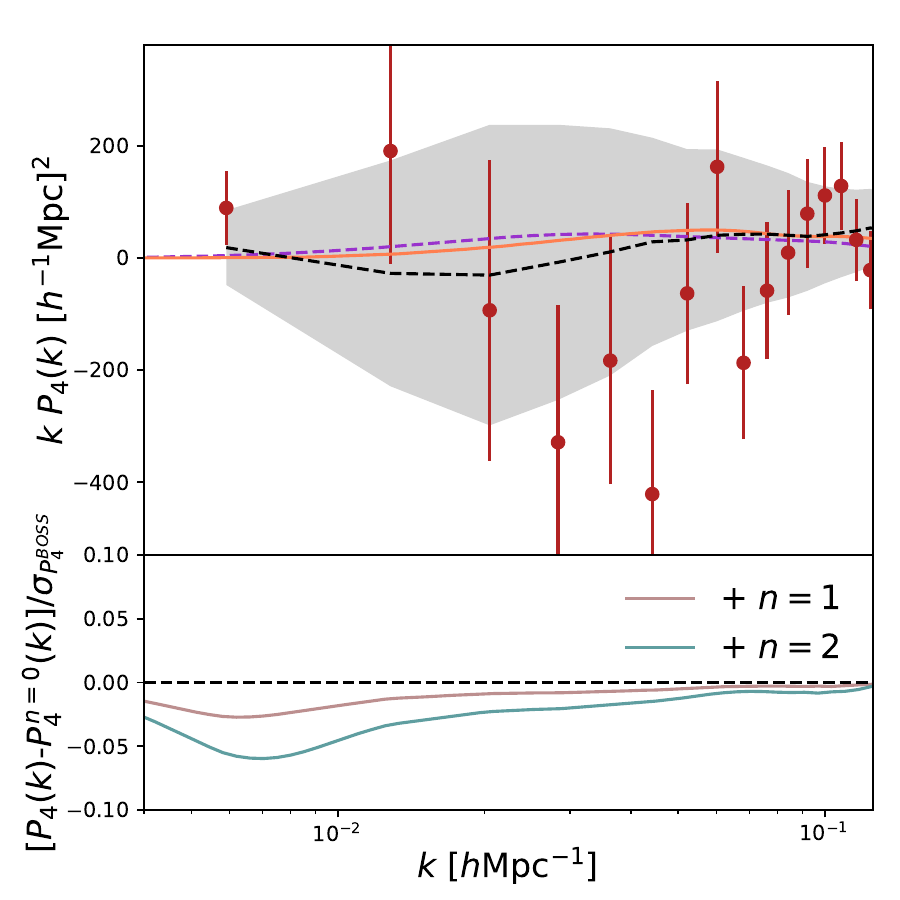}
\caption{The even power spectrum multipoles for the low redshift bin (z1) of BOSS DR12, NGC. The odd multipoles are shown in \fig{fig:measurement_odd_NGC}. The red data points show the BOSS DR12 measurements, while the black dashed line shows the mean of the Multidark Patchy mock catalogs, together with the variance around the mean (gray band). The orange solid line shows the best fitting model based on renormalized perturbation theory (TNS model), including all ($n < 3$) window function and wide-angle corrections. The fitting range is $0.008 < k < 0.096\kMpc$, indicated by the vertical gray dashed lines. The magenta dashed line shows the same best fitting model  without any wide-angle contribution (only $n=0$ terms). The lower panel shows the impact of the wide-angle corrections at order one ($n=1$) and order two ($n=2$) using the TNS model from the upper panel.}
\label{fig:measurement_even_NGC}
\end{figure}

\begin{figure}[t]
\centering
\includegraphics[width=0.485\textwidth]{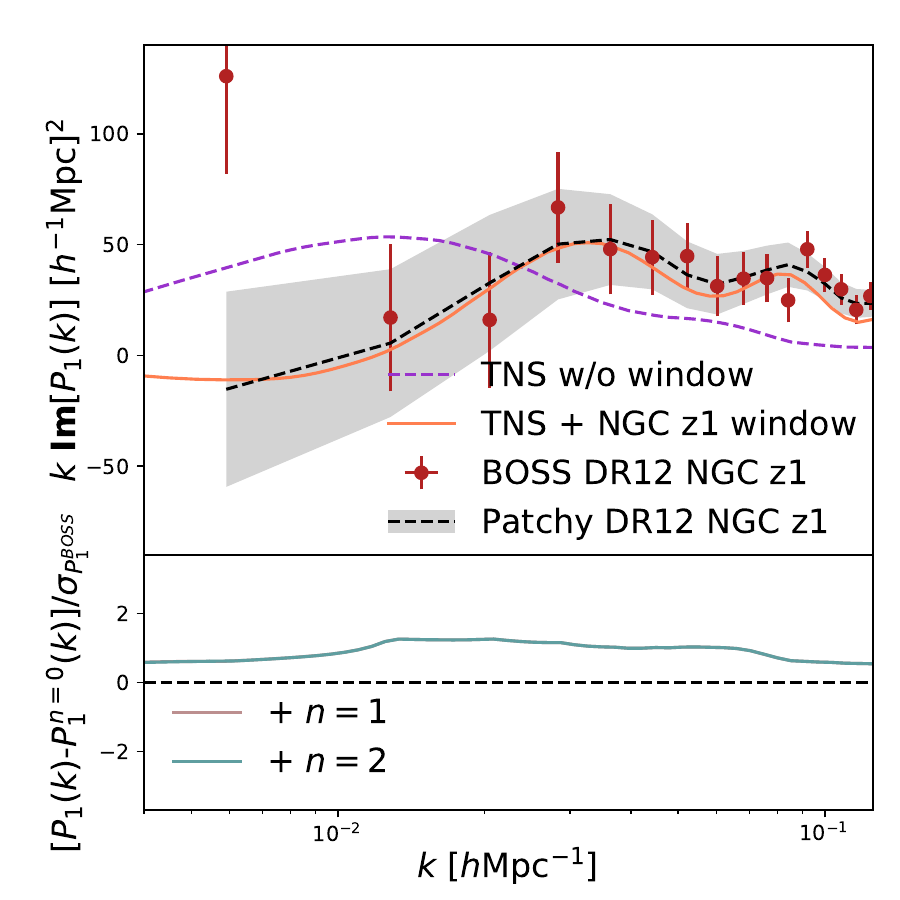}
\includegraphics[width=0.485\textwidth]{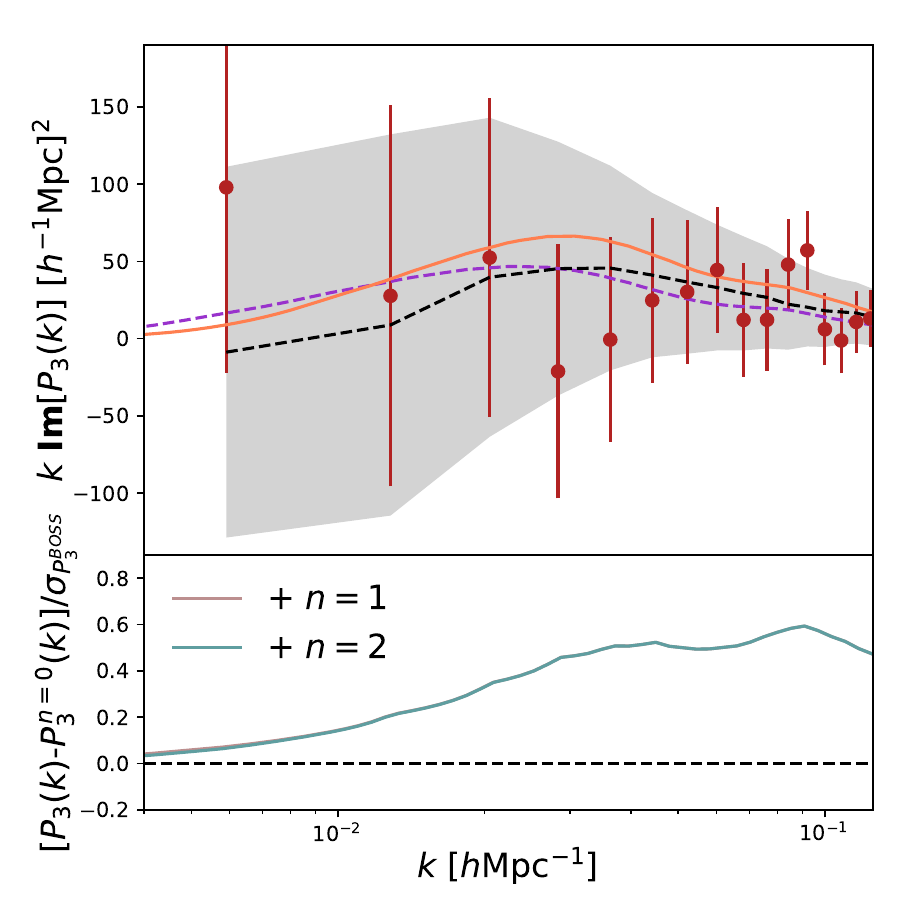}
\caption{The odd power spectrum multipoles for the low redshift bin (z1) of BOSS DR12, NGC. The even multipoles are shown in \fig{fig:measurement_even_NGC}. The red data points show the BOSS DR12 measurements, while the black dashed line shows the mean of the Multidark Patchy mock catalogs, together with the variance around the mean (gray band). The orange solid line shows the best fitting model based on renormalized perturbation theory (TNS model), including all ($n < 3$) window function and wide-angle corrections. The fitting range is $0.008 < k < 0.096\kMpc$, indicated by the vertical gray dashed lines. The magenta dashed line shows the same best fitting model  without any wide-angle contribution (only $n=0$ terms). The lower panel shows the impact of the wide-angle corrections at order one ($n=1$) and order two ($n=2$) using the TNS model from the upper panel. Be aware that computing the total significance of the dipole and octopole from only the lower panels shown above, neglects the strong correlation between different multipoles.}
\label{fig:measurement_odd_NGC}
\end{figure}

\begin{figure}[t]
\centering
\includegraphics[width=0.325\textwidth]{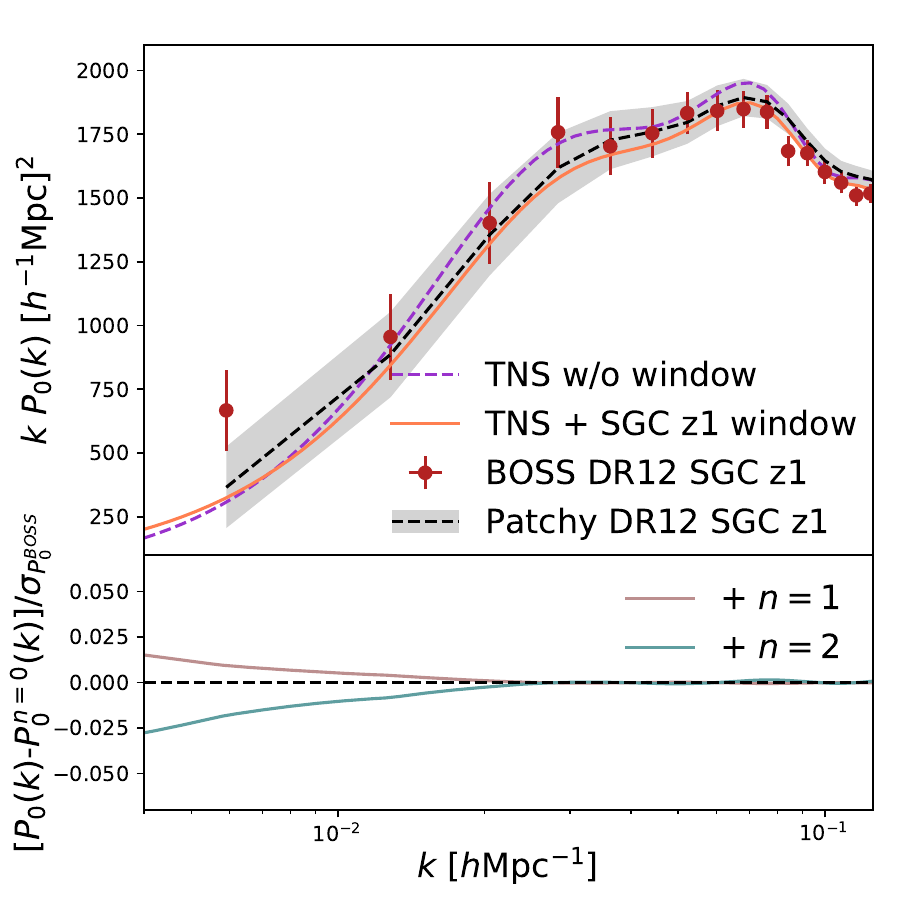}
\includegraphics[width=0.325\textwidth]{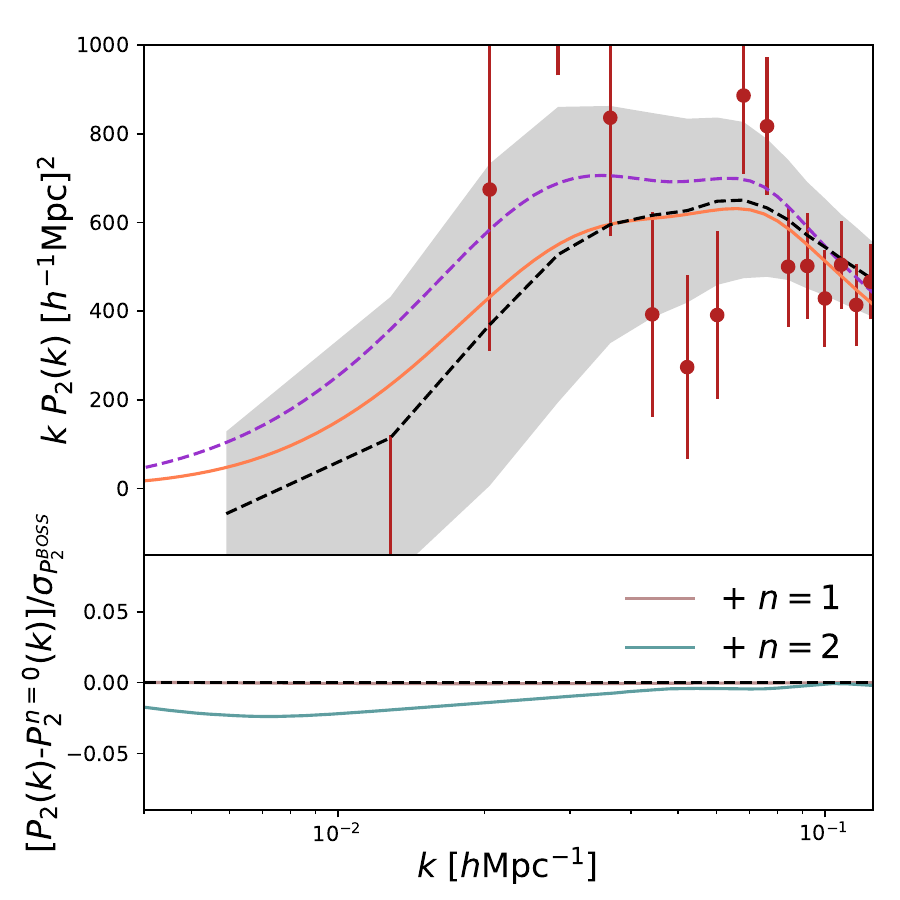}
\includegraphics[width=0.325\textwidth]{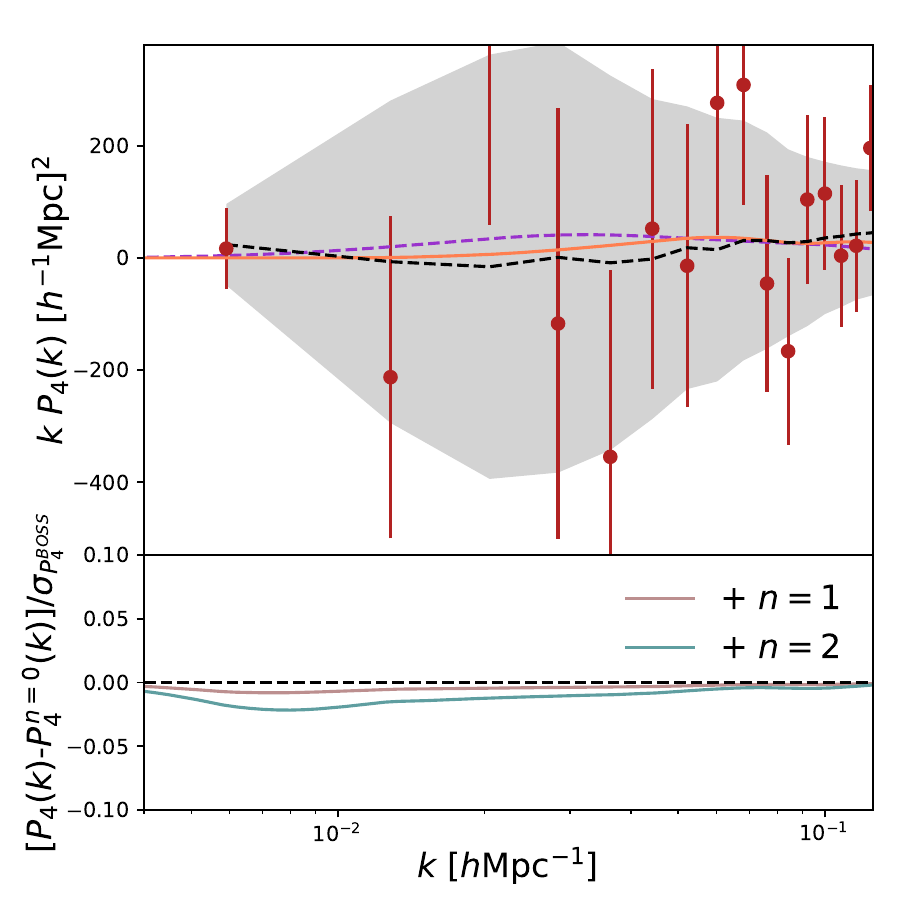}
\caption{The even power spectrum multipoles of BOSS DR12 SGC in the low redshift bin ($0.2 < z < 0.5$, $z_{\rm eff} = 0.38$). The equivalent measurements of the NGC are shown in \fig{fig:measurement_even_NGC}.}
\label{fig:measurement_even_SGC}
\end{figure}

\begin{figure}[t]
\centering
\includegraphics[width=0.485\textwidth]{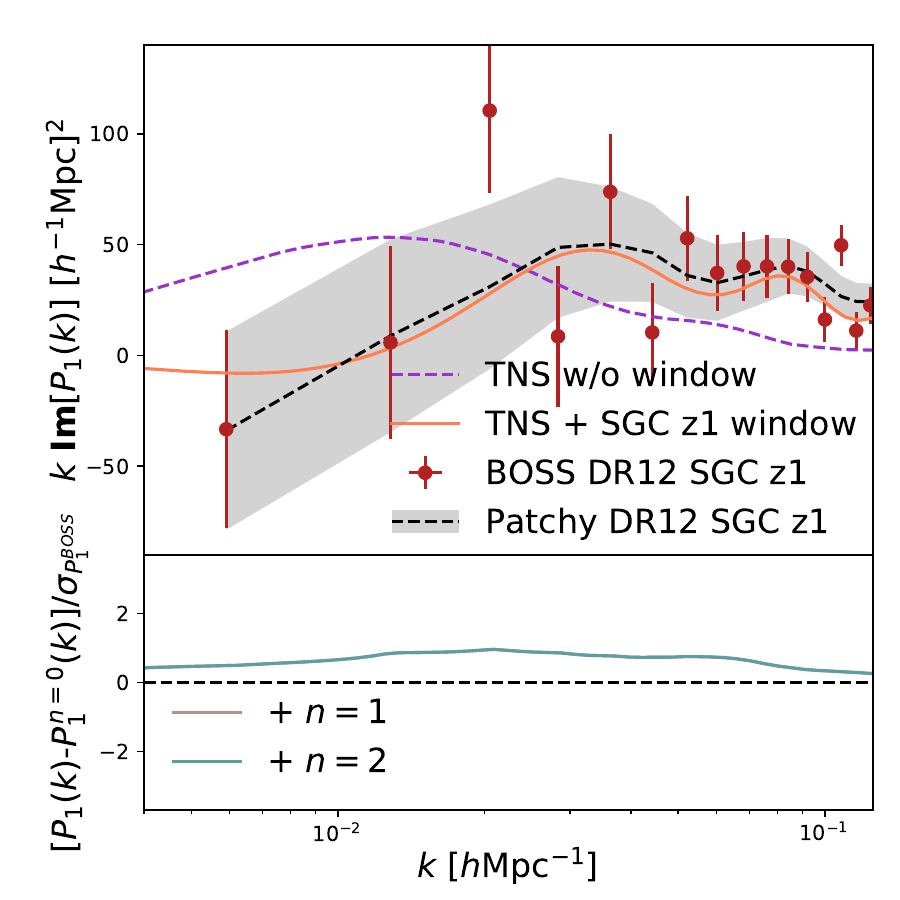}
\includegraphics[width=0.485\textwidth]{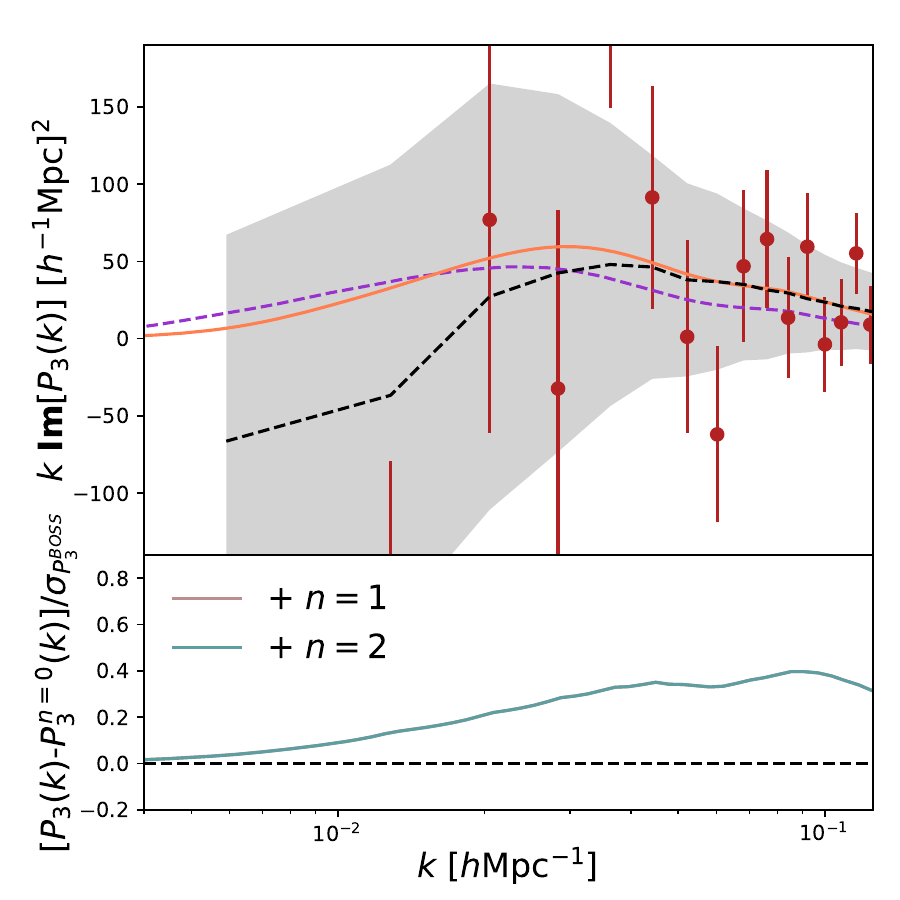}
\caption{The odd power spectrum multipoles of BOSS DR12 SGC in the low redshift bin ($0.2 < z < 0.5$, $z_{\rm eff} = 0.38$). The equivalent measurements of the NGC are shown in \fig{fig:measurement_odd_NGC}.}
\label{fig:measurement_odd_SGC}
\end{figure}

In the last sections we developed a formalism to account for wide-angle effects in the modeling of the power spectrum multipoles and discussed their importance for a correct estimate of the signal.
Now we will use these tools to measure the power spectrum multipoles in BOSS DR12, compute the detection significance of wide-angle effects, and any possible bias the latter could induce on cosmological parameters. We will start with the Multidark Patchy mock catalogs before moving on to the BOSS dataset itself.

\subsubsection{The mock catalogs}

The first step of the data analysis is to obtain the covariance matrix, reflecting the uncertainties and correlations of the power spectrum measurements. We obtain the covariance matrix from the $2048$ Multidark Patchy mock catalogs. \fig{fig:cov_NGC_z1} shows the correlation matrix
\begin{equation}
R_{ij} = \frac{C_{ij}}{\sqrt{C_{ii}C_{jj}}},
\end{equation}
derived from the covariance matrix $C_{ij}$ of all multipoles in the low redshift bin of BOSS DR12, NGC. As expected, there is a strong correlation between the even multipoles and the odd multipoles, as the latter are sourced by the former, see \eq{eq:order1term0} and~(\ref{eq:order1term1}). 

\fig{fig:measurement_even_NGC} and~\ref{fig:measurement_odd_NGC} show the mean of $2048$ power spectrum multipoles in the Multidark Patchy mock catalogs (black dashed line) together with the variance of the $2048$ realizations (gray shaded area). \fig{fig:measurement_odd_NGC} shows that the dipole power spectrum is clearly detected whereas measurement uncertainties in the octopole are still large. We will formally quantify the detection significance in Section~\ref{sec:sig}.

The lower panels of \fig{fig:measurement_even_NGC} and~\ref{fig:measurement_odd_NGC} show the impact of the wide-angle correction terms at order $n=1$ and $n=2$ relative to the BOSS uncertainties. As expected, most of the correction terms become largest on small $k$ (large scales), and contribute $< 1\%$ on scales larger than $k=0.1\kMpc$. For all of the even multipoles the wide-angle corrections are small compared to the error bars, $5\%$ at most on large scales, while they significantly impact the odd multipoles.

We can now ask the question whether any cosmological parameter could be biased when the odd multipoles are ignored in the parameter fit. We use the covariance matrix based on the Patchy mock catalogs and analyze the mean of the $2048$ Patchy power spectrum multipoles in the $k$-range $0.008 < k < 0.096\kMpc$ in bins of $\Delta k = 0.008\kMpc$. We chose this binning since it is larger than the fundamental mode of the survey as shown in \fig{fig:kwindow} and a significantly smaller bin size would lead to correlated bins. We use a model for the anisotropic power spectrum based on renormalized perturbation theory (TNS model, \citep{Taruya2010:1006.0699v1}). A detailed description of this model and its implementation can be found in~\citep{Beutler2013:1312.4611v2,Beutler2016:1607.03150v1}. We follow the procedure used in these previous papers, meaning we fit the NGC and SGC simultaneously, using separate nuisance parameters for the two sky patches but common cosmological parameters.

Comparing the fit to all $5$ multipoles with the fit to only the even multipoles results in $\Delta\alpha_{\perp} = 0.00031$, $\Delta\alpha_{\parallel} = 0.0025$ and $\Delta f\sigma_{8} = 0.0059$, significantly below the BOSS measurement uncertainties. We therefore conclude that wide-angle effects do not impact the RSD and BAO constraints of BOSS DR12. 

However, the biggest impact of the wide-angle effects is on the largest scales, which are used to constrain e.g. primordial non-Gaussianity. It should be clear that the careful treatment of the survey window function and wide-angle effects is essential to avoid biases when using large-scale modes of datasets like BOSS. 

\subsubsection{The dataset}
\label{sec:dataanalysis}

We now move to the actual BOSS DR12 dataset. We use the covariance matrix introduced in the last section, to fit all $5$ power spectrum multipoles in the $k$-range $0.008 < k < 0.096\kMpc$ in bins of $\Delta k = 0.008\kMpc$ again using the TNS model. Here we are not aiming to model the data down to the smallest scales possible, but rather want to demonstrate the effectiveness of the formalism presented in this paper.

Fitting the low redshift bin of BOSS DR12 results in $\chi^2/{\rm d.o.f.} = 78.4/(110 - 11)$. Our fit has $110$ data bins ($11$ bins per multipole with $5$ multipoles per Galactic Cap) and $11$ free fitting parameters. The reduced $\chi^2$ indicates a good fit to the data and the cosmological parameters are in agreement with~\cite{Beutler2013:1312.4611v2,Beutler2016:1607.03150v1} and are given by $\alpha_{\perp} = 0.997\pm 0.066$, $\alpha_{\parallel} = 0.993\pm 0.075$ and $f\sigma_8 = 0.484\pm 0.062$. The best fitting model is included in \fig{fig:measurement_even_NGC} - \ref{fig:measurement_odd_SGC} (solid orange line). The high redshift bin yields very similar results with $\chi^2/{\rm d.o.f.} = 75.3/(110 - 11)$. We performed similar fits with a linear Kaiser model, but we found the TNS model is a better description of the quadrupole even on these very large scales (as previously shown in e.g.~\citep{Torre2012:1202.5559v4}).

\subsubsection{Detection significance}
\label{sec:sig}

To determine the detection significance of the odd multipoles we set the model for the odd multipoles to zero and refit the data. For the low redshift bin of BOSS DR12 this results in $\chi^2/{\rm d.o.f.} = 159.5/(110 - 11)$, indicating that this model is a bad description of the data. Comparing with the original $\chi^2=78.4$, we obtain a detection significance for the odd multipoles of $\sqrt{\Delta\chi^2} = 9.0\sigma$. For the high redshift bin we get $\Delta\chi^2 = 150.5 - 75.3 = 75.2$, resulting in a $8.7\sigma$ detection of the odd multipoles.
It is worth pointing out that the detection of the BAO in the BOSS data is around $8\sigma$, and therefore comparable to the detection of the dipole and the octopole.

If we exclude the octopole from the exercise above we get a detection significance for the dipole alone of 
%65.9 - 125.8
$7.7\sigma$ for the low redshift bin and 
%71.3 - 115.3
$6.6\sigma$ for the high redshift bin. If we remove the dipole, the detection significance for the octopole in the low redshift bin is
% 68.5 - 87.1
$4.3\sigma$
and in the high redshift bin we get
% 67.6 - 85.3
$4.2\sigma$.

Next we want to determine the significance of the $n>0$ contributions. We fit all $5$ power spectrum multipoles but now we set all $n>0$ terms to zero. This keeps the standard window function terms at $n=0$, but removes all wide-angle correction terms. In this case we get $\chi^2/{\rm d.o.f.} = 114.2/(110 - 11)$ indicating a detection significance of the wide-angle terms of $6.0\sigma$. For the high redshift bin we get $\chi^2/{\rm d.o.f.} = 115.9/(110 - 11)$ resulting in a $6.4\sigma$ significance.

The values reported in this section are dependent on the chosen value of $k_{\rm max}$ and we did not use the small scale regime as much as we could have. 
Also, be aware that judging the significance of the dipole and octopole, by eye, from the lower panels of \fig{fig:measurement_odd_NGC} and~\ref{fig:measurement_odd_SGC}, neglects the strong correlation between different multipoles (see \fig{fig:cov_NGC_z1}).
The main message of this section is that the BOSS dataset allows a clear detection of the wide-angle effects.

\section{Conclusion}
\label{sec:conclusion}

In this work we presented a formalism to account for the wide-angle effects in the power spectrum multipoles and their interplay with the survey window function. While wide-angle effects are present in all 2-point function estimators, they are more prominent in FFT-based power spectrum estimators, which rely on the end-point line-of-sight definition. Based on~\citep{Castorina2017:1709.09730v2,Castorina2018:1803.08185v2}, we propose a formalism to include wide-angle effects self-consistently in the power spectrum model, by expanding the usual treatment of the survey window function.

We present all wide-angle correction terms up to second order ($n < 3$) for the high and low redshift bins of BOSS DR12. We test our formalism on the Multidark Patchy mock catalogs as well as the BOSS data. We found that the survey window function and wide-angle correction terms can contribute up to $5\%$ of the measurement uncertainties in the even multipoles, while odd power spectrum multipoles are completely dominated by the wide-angle correction terms on all scales. 

Using a model based on renormalized perturbation theory (TNS model) we are able to fit all $5$ power spectrum multipoles in the $k$-range $0.008 < k < 0.096\kMpc$, resulting in $\chi^2/{\rm d.o.f.} = 78.4/(110 - 11)$ for the low redshift bin and $\chi^2/{\rm d.o.f.} = 75.3/(110 - 11)$ for the high redshift bin, indicating that our model is a good description for the data. We clearly detect the non-zero odd power spectrum multipoles in both redshift bins with $\sim9\sigma$ significance. 

The odd power spectrum multipoles are generated by leakage of power from the even multipoles as well as wide-angle effects at order $n=1$. We detect the wide-angle component with $>6\sigma$ in both redshift bins. The good fit of the TNS model to the data indicates that our measurements of the odd multipoles are consistent with being sourced by the window function and wide-angle effects.

Including the wide-angle correction terms is essential for many cosmological observables, like e.g. primordial non-Gaussianity, which relies on accurate models of the low-k modes or the search for GR effects, which tries to detect a signal in the odd multipoles. While in the case of BOSS we could not find a significant impact of wide-angle effects on measurements of RSD or BAO, future surveys like DESI and Euclid will require much higher accuracy of the clustering models. Our formalism is well suited to capture wide-angle effects and hence should alleviate the potential biases due to FFT-based power spectrum estimators.

\section*{Acknowledgments}

FB would like to thank Chris Blake, Michael Wilson and Morag Scrimgeour for comments on an early draft of this paper and EC would like to thank Martin White and Enea di Dio for very useful discussions on wide-angle effects.
FB is a Royal Society University Research Fellow.

\bibliographystyle{JHEP}
\bibliography{main}{}

\newpage

\appendix

\section{Derivation of \eq{eq:convolvedP}}

We start with 
\begin{equation}
\avg{\convolved{P}_A(\vb{k})} = (2A+1)\int \frac{d \Omega_k}{4\pi} \,\mathrm{d}^3 s_1 \,\mathrm{d}^3 s_2 \,e^{-i \vb{k}\cdot(\vb{s}_1-\vb{s}_2)} \avg{\delta(\vb{s}_1) \delta(\vb{s}_2)}  W(\vb{s}_1)W(\vb{s}_2)\mathcal{L}_A(\vbunit{k}\cdot \vbunit{s}_1)
\end{equation}
and use 
\begin{align}
e^{-i\vb{k}\cdot \vb{r}} &= \sum_s(-i)^s(2s+1)j_s(kr)L_s(\vbunit{k}\cdot\vbunit{r})\\
\avg{\delta(\vb{s}_1) \delta(\vb{s}_2)} &= \sum_{n,\ell} x_s^n\xi^n_{\ell}(s)\mathcal{L}_{\ell}(\vbunit{s}\cdot\vbunit{s}_1)\\
(2\ell + 1)\int \frac{d\Omega_k}{4\pi}\mathcal{L}_{\ell}(\vbunit{k}\cdot\vbunit{s})\mathcal{L}_{\ell'}(\vbunit{k}\cdot\vbunit{s}) &= \mathcal{L}_{\ell}(\vbunit{s}_1\cdot\vbunit{s})\delta_{\ell\ell'}
\end{align}
to get 
\begin{equation}
\avg{\convolved{P}_A(\vb{k})} = (2A+1)\int \mathrm{d}^3 s_1 \,\mathrm{d}^3 s_2 (-i)^A j_A(ks) \mathcal{L}_{A}(\vbunit{s}_1\cdot\vbunit{s}) \sum_{n,\ell}x_s^n\xi^{(n)}_{\ell}(s)\mathcal{L}_{\ell}(\vbunit{s}_1\cdot\vbunit{s})W(\vb{s}_1)W(\vb{s}_2)\mathcal{L}_A(\vbunit{k}\cdot \vbunit{s}_1).
\end{equation}
We now use 
\begin{equation}
\mathcal{L}_{\ell}(\vbunit{s}_1\cdot\vbunit{s})\mathcal{L}_{A}(\vbunit{s}_1\cdot\vbunit{s}) = \sum_L \tj{\ell}{L}{A}{0}{0}{0}^2(2L + 1)\mathcal{L}_{L}(\vbunit{s}_1\cdot\vbunit{s})
\end{equation}
to get to \eq{eq:convolvedP}
\begin{equation}
	\begin{split}
		\avg{\convolved{P}_A(\vb{k})} &= (-i)^A (2A+1) \sum_{\ell,\, L}\tj{\ell}{L}{A}{0}{0}{0}^2(2L+1)\int \mathrm{d}s\,s^2 j_A (ks) \sum_n (s)^n\, \xi_\ell^{{\rm ep},(n)}(s) \\
		&\times \int \mathrm{d} \Omega_s \int \mathrm{d}^3 s_1  (s_1)^{-n} W(\vb{s}_1)W(\vb{s}+\vb{s}_1)\mathcal{L}_L(\vbunit{s}\cdot \vbunit{s}_1),
	\end{split}
\end{equation}
where we also used $x^n_s = \frac{s^n}{d^n}$ with $d = |\vb{s}_1|$ in the case of the end-point LOS.

\section{Selection function terms}

\label{sec:selfunc}
The evolution of the mean number density of galaxies with redshift, \ie distance to the observer, can generate additional wide-angle contributions~\cite{Kaiser,Szalay1997:astro-ph/9712007v1,Hamilton1997:astro-ph/9708102v2,PapSza08,Castorina2018:1803.08185v2}\footnote{See~\cite{Yoo} for the connection of these terms to a fully relativistic treatment.}.  
The size of these selection function terms is proportional to
\begin{align}
\label{eq:alpha}
\alpha(\vb{r}) = \frac{\mathrm{d} \log r^2 \bar{n}_g (\vb{r})}{\mathrm{d} \log r } = 2 + \frac{r H(z)}{1+z}\frac{\mathrm{d} \log \bar{n}_g (z)}{\mathrm{d} \log (1+z) }\equiv 2 + \frac{r H(z)}{1+z} \gamma(z)
\end{align}
with $\vb{r}$ being the \emph{real space} distance vector to redshift $z$ and $\bar{n}_g (\vb{r})$ the galaxy density at that distance. The first term on the right hand side corresponds to a redshift independent selection function, whereas the second one accounts for the redshift evolution of the sample,  and for later convenience we defined $\gamma(z) = \frac{\mathrm{d} \log \bar{n}_g (z)}{\mathrm{d} \log (1+z) }$.
In a galaxy survey one has typically access to the redshift space number density of galaxies, $\bar{n}_g(\vb{s})$, which should therefore be converted back to real space in order to be able to compute the selection function contribution to the galaxy power spectrum multipoles. This step can be quite complicated in practice \cite{Hamilton1997:astro-ph/9708102v2}.
As shown in~\citep{Castorina2017:1709.09730v2}, if the selection function is constant in redshift the new wide-angle contributions up to  $\mathcal{O}(x_s^2)$ are
\begin{align}
\label{eq:xialpha}
\xi_0(s, d) &\ni  b^2 \frac{4\beta^2}{3 d^2}\ \Xi_0^{(2)}(s)
  + \frac{2}{3 d} b^2 \beta(1-\beta)x_s\ \Xi_1^{(1)}(s) \notag \\
 \xi_2(s, d) &\ni  -\frac{8}{3 d^2}b^2 \beta^2\ \Xi_2^{(2)}(s)
     -\frac{8}{15 d}b^2 \beta(5+\beta)x_s\ \Xi_1^{(1)}(s) +\frac{4}{5 d} b^2 \beta^2x_s\ \Xi_3^{(1)}(s)
\end{align}
and no effect on the hexadecapole.

The second term on the right hand side of \eq{eq:alpha} is qualitatively different from other wide-angle effects, since it is suppressed by the physical scale associated with the evolution of the galaxy sample, \ie the Hubble rate $H(z)$, rather than by the distance $\vb{d}$ of the observer to the galaxies.
At next to leading order the relevant contributions to the multipole correlation function are
\begin{align}
\xi_0(s, d) &\ni \xi^{{\rm sel},(\gamma)}_0(s) = -\frac{2}{15}b^2 \beta(\beta-5) \frac{H(z)}{1+z}\gamma(z) \,x_s \Xi_1^{(1)}(s) + \frac{1}{3}b^2 \beta^2\left[\frac{H(z)}{1+z}\gamma(z)\right]^2 \Xi_0^{(2)}(s)  \notag \\
 \xi_2(s, d) &\ni \xi^{{\rm sel},(\gamma)}_2(s) = \frac{2}{105}b^2 \beta \frac{H(z)}{1+z}\gamma(z)x_s[12 \beta \Xi_3^{(1)}(s)+7(\beta-5)\Xi_1^{(1)}(s)]-\frac{2}{3} b^2 \beta^2 \left[\frac{H(z)}{1+z}\gamma(z)\right]^2 \Xi_2^{(2)}(s) \notag \\
 \xi_4(s, d) &\ni \xi^{{\rm sel},(\gamma)}_4(s) = -\frac{8}{35}b^2 \beta^2 \left[\frac{H(z)}{1+z}\gamma(z)\right]^2 \Xi_3^{(1)}(s).
\end{align}
As discussed above, these terms are suppressed by powers of $k(a H)^{-1}$, and are therefore sub-dominant if $d\lesssim H^{-1}$, as it is the case for the BOSS survey. They can also be integrated by parts in redshift to zero if the selection function is negligible at the two extrema of the redshift distribution \cite{Szalay1997:astro-ph/9712007v1}.
Given the difficulty in estimating $\gamma(z)$ from the data, we do not include the selection function terms in our analysis. 
Reference \citep{Castorina2017:1709.09730v2} showed that selection function terms are small compared to the wide-angle effects introduced by the choice of an asymmetric LOS (see also \cite{Raccanelli2010:1006.1652v1,Yoo2015} for a discussion of these terms when $\alpha(r)$ is known).

\section{Power spectrum wedges}
\label{sec:wedges}

\begin{figure}
\centering
\includegraphics[width=0.5\textwidth]{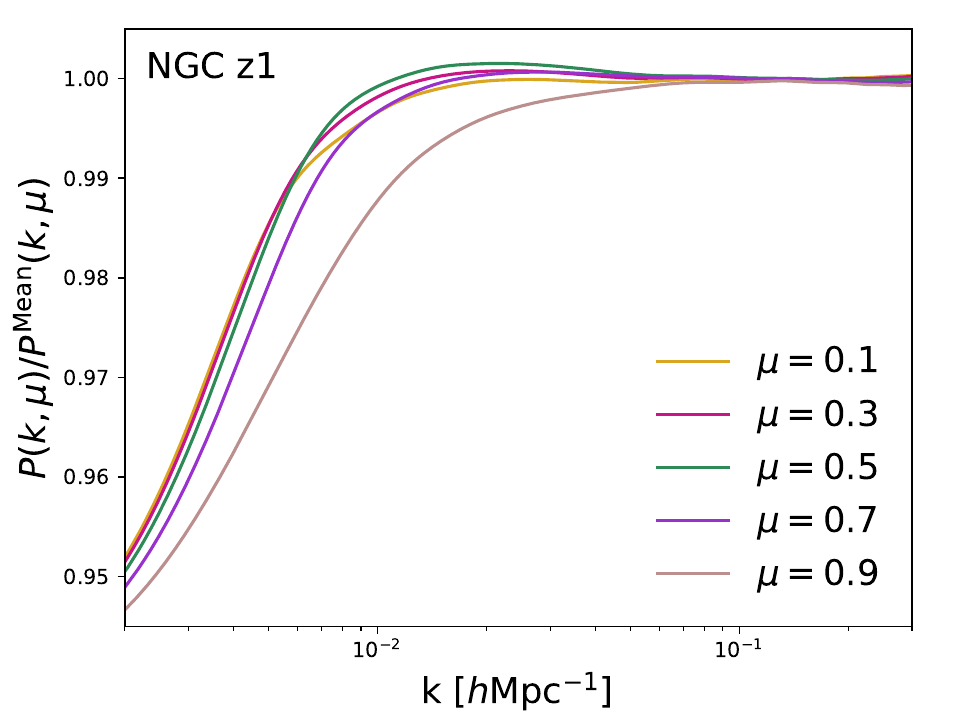}
\caption{This plot shows the effect of using inconsistent LOS choices for the power spectrum estimate and the window function estimate on the power spectrum wedges. This Figure can be compared to \fig{fig:old_new_cmp} in Section~\ref{sec:formerwork}, which shows the effect on the multipoles. The y-axis shows the ratio between the power spectrum wedges including all wide-angle corrections consistently using the end-point LOS, relative to the case where the convolved power spectrum wedges have been calculated using the mean LOS window function at $n=0$ (consistent with the treatment in~\citep{Grieb}). 
Here we used the low redshift bin of BOSS DR12 NGC with $\Delta \mu = 0.2$.
The differences reach up to $5\%$ at the smallest $k$ and are more important for high $\mu$.}
\label{fig:wedges}
\end{figure}

In the plane-parallel limit, or in the flat sky case, the redshift space clustering can be interchangeably described by multipoles of the power spectrum or by binning the two-dimensional power spectrum in wedges of $\mu$.
In a curved sky, however, there is no unique definition of the parallel, $k_{||}$, and transverse, $k_{\perp}$, components of the 3-dimensional wave-vector $\vb{k}$. This is just a consequence of the fact that the line of sight varies across the survey. We can nonetheless construct a wedge-like summary statistic by summing the FFT-estimated multipoles with appropriate weighting \cite{Grieb}
\begin{align}
\convolved{P}(k,\mu) \equiv \sum_\ell \convolved{P}_\ell(k) \bar{\mathcal{L}}_\ell(\mu)
\label{eq:Pkwedges}
\end{align}
with $\bar{\mathcal{L}}_\ell(\mu)$ being the average value of the Legendre polynomial
\begin{align}
\bar{\mathcal{L}}_\ell(\mu) \equiv \frac{1}{\Delta \mu} \int_{\mu -\Delta \mu /2}^{\mu + \Delta \mu /2} \rm{d} \nu \,\mathcal{L}_\ell(\nu) 
\end{align}
for some binning $\Delta \mu$. It is easy to convince ourselves that the estimator in \eq{eq:Pkwedges} has the appropriate plane-parallel limit and it is therefore suitable for cosmological parameter inference.

The clustering wedges defined in this way inherit the wide-angles structure of the power spectrum multipoles discussed in Section~\ref{sec:wideangle}. An inconsistent treatment of the LOS in the power spectrum and window function will lead to $n=0$ biases, whereas intrinsic wide-angle terms will contribute at $n=1,2$. \fig{fig:wedges} shows the impact of inconsistent LOS choices on the clustering wedges for the low redshift bin of BOSS NGC, in bins of $\Delta \mu = 0.2$. The numerator of the quantity plotted on the y-axis represents the wedges computed with a consistent LOS treatment up to $n=2$ (using the end-point LOS). The denominator used the end-point LOS for the power spectrum and the mean LOS for the window function, only including the terms in the plane-parallel limit, $n=0$. This plot is analog to the lower panels in \fig{fig:old_new_cmp} for the power spectrum multipoles.
To define the wedges we include all multipoles up to $\ell=4$, neglecting higher order ones that could be important at high $k$. This assumption is justified as wide-angle affects are only important at large scales.
Our results indicate that ignoring wide-angle effects could potentially bias the estimated wedges by at most $5\%$ for $k\le 0.01 \kMpc$, with increasing error for larger values of $\mu$.
We, however, expect wide-angle effect to play a minor role in the analysis of~\cite{Grieb} given the current measurement uncertainties of BOSS.

\section{Explicit expression for the convolved even multipoles}
\label{App:even}

The power spectrum monopole, including the convolution with the survey window function is given by
\begin{align}
	\convolved{P}_0(k) &= \sum_{\ell,\, L}\tj{\ell}{L}{0}{0}{0}{0}^2\int \mathrm{d}s\ j_0 (ks) \sum_n s^{n+2}\xi_\ell^{(n)}(s)  Q_L^{(n)}(s) \\ &=\sum_n \int ds\, j_{0}(ks)s^{n+2} \convolved{\xi}_{0}^{(n)}(s)\,,
\end{align}
The convolved correlation function at order $n=0$ and $n=2$ have the same general form and are given by
\begin{align}
	\convolved{\xi}_0^{(n)}(s) = \xi_0^{(n)}(s)Q_0^{(n)}(s)+\frac{1}{5}\xi_2^{(n)}(s)Q_2^{(n)}(s)+\frac{1}{9}\xi_4^{(n)}(s)Q_4^{(n)}(s)+\dots,
\end{align}
which for $n = 0$ recovers the standard expressions of~\citep{Wilson2015:1511.07799v2} and~\citep{Beutler2016:1607.03150v1}. 
The dipole and the octopole of the underlying galaxy distribution contribute to the monopole via coupling with the window function, and start at $n=1$,
\begin{align}
	\convolved{\xi}_0^{(1)}(s) = \frac{1}{3}\xi_1^{(1)}(s)Q_1^{(1)}(s)+\frac{1}{7}\xi_3^{(1)}(s)Q_3^{(1)}(s) + \cdots.
\end{align}
We proceed similarly for the higher order multipoles.
The convolved correlation function quadrupole and hexadecapole at $n=0,2$ are given by
\begin{align}
	\begin{split}
		\convolved{\xi}_2^{(n)}(s) = \xi_0^{(n)}(s)Q^{(n)}_{2}(s) &+ \xi_2^{(n)}(s)\left[Q^{(n)}_0(s) + \frac{2}{7}Q^{(n)}_2(s) + \frac{2}{7}Q^{(n)}_4(s)\right]\\
		&+\xi_4^{(n)}(s)\left[\frac{2}{7}Q^{(n)}_2(s) + \frac{100}{693}Q^{(n)}_4(s) + \frac{25}{143}Q^{(n)}_6(s)\right]\\
		&+\cdots
	\end{split}\\
	\label{eq:std_quad}
	\begin{split}
		\convolved{\xi}_4^{(n)}(s) = \xi_0^{(n)}(s)Q^{(n)}_{4}(s) &+ \xi_2^{(n)}(s)\left[\frac{18}{35}Q^{(n)}_2(s) + \frac{20}{77}Q^{(n)}_4(s) + \frac{45}{143}Q^{(n)}_6(s)\right]\\
		&+\xi_4^{(n)}(s)\Bigg[Q^{(n)}_0(s) + \frac{20}{77}Q^{(n)}_2(s) + \frac{162}{1001}Q^{(n)}_4(s) + \frac{20}{143}Q^{(n)}_6(s) +\frac{490}{2431}Q^{(n)}_8(s)\Bigg]\\
		&+\cdots
	\end{split}
\end{align}
The leading order ($n$=1) corrections instead are
\begin{align}
	\begin{split}
		\convolved{\xi}_2^{(1)}(s) &= \xi_1^{(1)}(s)\left[\frac{2}{3}Q_1^{(1)}(s)+\frac{3}{7} Q_3^{(1)}(s)\right]\\
        &+ \xi_3^{(1)}(s)\left[\frac{3}{7}Q_1^{(1)}(s)+\frac{4}{21} Q_3^{(1)}(s)\right]\\
        & + \cdots
	\end{split}
\end{align}
and
\begin{align}
	\begin{split}
		\convolved{\xi}_4^{(1)}(s) = \frac{4}{7}\xi_1^{(1)}(s)Q_3^{(1)}(s) + \xi_3^{(1)}(s)\left[\frac{4}{7}Q_1^{(1)}(s)+\frac{18}{77} Q_3^{(1)}(s)\right] + \cdots
    \end{split}
\end{align}

\section{FFT-based dipole and octopole estimator}
\label{app:estimators}

Here we present the estimators for the first two odd power spectrum multipoles in the Legendre basis following the nomenclature of~\citep{Bianchi2015:1505.05341v2}. The estimators are defined by
\begin{align}
P_1(k) &= \frac{3}{I}\int\frac{d\Omega_k}{4\pi}A_0(\vb{k})A^*_{1}(\vb{k})\label{eq:dipole}\\
P_3(k) &= \frac{7}{2I}\int\frac{d\Omega_k}{4\pi}A_0(\vb{k})\left[5A^*_3(\vb{k}) - 3A^*_1(\vb{k})\right]
\label{eq:octopole}
\end{align}
with
\begin{equation}
A_{\ell}(\vb{k}) = \int d\vb{r}\;(\vbunit{k}\cdot \vbunit{r})^{\ell}F(\vb{r})e^{i\vb{k}\cdot\vb{r}},
\label{eq:overdensity}
\end{equation}
and
\begin{equation}
    F(\vb{r}) = w(\vb{r})\left[n_{\rm gal}(\vb{r}) - \alpha n_{\rm ran}(\vb{r})\right].
\end{equation}
The normalization is given by
\begin{align}
    I &= \int d\vb{r}\; w^2(\vb{r})n_{\rm gal}^2(\vb{r})\\
    &= \sum^{N_{\rm ran}}_iw^2(\vb{r}_i)n_{\rm gal}(\vb{r}_i),
\end{align}
where $\vb{r}$ is the 3D position of a grid cell of the binned density field and $\vb{r}_i$ is the position of the $i^{\rm th}$ galaxy. The dipole and octopole terms of $A_{\ell}$ are given by
\begin{align}
  A_1(\vb{k}) &= \frac{1}{k}\left(k_xD_x(\vb{k}) + k_yD_y(\vb{k}) + k_zD_z(\vb{k})\right),\\
  \begin{split}
    A_3(\vb{k}) &= \frac{1}{k^3}\Bigg(k^3_xQ_{xxx}(\vb{k}) + k^3_yQ_{yyy}(\vb{k}) + k^3_{z}Q_{zzz}(\vb{k})\\
    &+ 3\Big[k^2_xk_yQ_{xxy}(\vb{k}) + k_x^2k_zQ_{xxz}(\vb{k}) + k_y^2k_xQ_{yyx}(\vb{k}) +\\ 
    &\;\;\;\;\;\;\;\;k_y^2k_zQ_{yyz}(\vb{k}) + k_z^2k_xQ_{zzx}(\vb{k}) + k_z^2k_yQ_{zzy}(\vb{k})\Big]\\
    &+ 6k_xk_yk_zQ_{xyz}(\vb{k})\Bigg)
  \end{split}
\end{align}
with
\begin{align}
D_i(\vb{k}) &= \int d\vb{r}\;\frac{r_i}{r}F(\vb{r})e^{i\vb{k}\cdot \vb{r}},\\
Q_{ijk}(\vb{k}) &= \int d\vb{r}\;\frac{r_{i}r_{j}r_{k}}{r^3}F(\vb{r})e^{i\vb{k}\cdot \vb{r}}.
\end{align}
The expression in \eq{eq:dipole} and~(\ref{eq:octopole}) are complex quantities with zero real part, since the real terms are anti-symmetric in $\vb{k}$ and therefore average to zero. This leads to
\begin{align}
P_1(k) = \frac{3}{I}\int\frac{d\Omega_k}{4\pi}&\left[A_0(\vb{k})A^*_1(\vb{k}) \right] \notag \\
 =  \frac{3}{I}\int\frac{d\Omega_k}{4\pi}&\Big\{\mathrm{Re} [A_0(\vb{k})] \mathrm{Re} [A_1(\vb{k})] +\mathrm{Im} [A_0(\vb{k})] \mathrm{Im} [A_1(\vb{k})]  \notag \\
&+ i \,\mathrm{Im} [A_0(\vb{k})] \mathrm{Re} [A_1(\vb{k})] - i\,\mathrm{Re} [A_0(\vb{k})] \mathrm{Im} [A_1(\vb{k})]\Big\} \notag \\ 
=\frac{3i}{I}\int\frac{d\Omega_k}{4\pi}& \Big\{\mathrm{Im} [A_0(\vb{k})] \mathrm{Re} [A_1(\vb{k})] - \mathrm{Re} [A_0(\vb{k})] \mathrm{Im} [A_1(\vb{k})] \Big \}\\
P_3(k) = \frac{7i}{2I}\int\frac{d\Omega_k}{4\pi}& \Big\{\mathrm{Im} [A_0(\vb{k})] \mathrm{Re} [5A_3(\vb{k}) - 3A_1(\vb{k})]  - \mathrm{Re} [A_0(\vb{k})]\mathrm{Im} [5A_3(\vb{k}) - 3A_1(\vb{k})] \Big \}.
\end{align}
The same multipoles can also be measured with the spherical harmonic basis following~\citep{Hand2017:1704.02357v1}, and we intent to include such estimators in \texttt{nbodykit}~\citep{Hand2017:1712.05834v1} in due course.

\subsection{Discussion on FFT-based window function estimation}
\label{sec:FFTwindow}

The estimators discussed above are not just useful to estimate the power spectrum multipoles, but they can also be used to estimate the window function multipoles. We just have to replace the over-density field $F(\vb{r})$ in \eq{eq:overdensity} with the density in the random catalog. Figure~\ref{fig:kwindow} shows the multipoles of the survey window in Fourier-space for the low redshift bin of BOSS DR12. This window function has been estimated in a cubic box with a side-length of $100\Gpc$. The large volume allows a better sampling of the largest modes, but there is still some noise visible, especially in the hexadecapole component. 

The window function derived using the FFT-based estimator has a Nyquist frequency, introduced by the binning of the density field, just as present in the power spectrum. On scales below the Nyquist frequency the window is smoothed out. 

To obtain the configuration-space window function required for the corrections in Section~\ref{sec:wideangle}, we have to Fourier-transform the k-space window function using an inverse Hankel transform. Such a transform puts extra strain on the analysis, since it can lead to oscillations at large scales due to the lack of sampling in Fourier-space. Given that the higher order window function terms are multiplied by $\chi^n$, such issues would be enhanced for higher order corrections. Fitting polynomials to the k-space window before the Hankel transform or smoothing the Fourier-transformed window function can mitigate such effects.

We therefore conclude that the FFT-based estimator is the natural choice for any Fourier-space analysis, since it offers the same computational complexity as the power spectrum estimator, while the Nyquist frequency limitations are similar to the situation in the power spectrum. We note however, that the alternative pair counting analysis does not necessarily require a large number of galaxies $N$ and hence even a $\mathcal{O}(N^2)$ analysis is feasible. However, the noise on small scales does often require a smoothing procedure or a cut at small scales, which has a similar effect as the Nyquist frequency. 

\end{document}